\documentclass[usenatbib]{aa}

\usepackage{comment}
\usepackage{placeins}
\usepackage{graphicx}	
\usepackage{amsmath,bm}	
\usepackage{wasysym}    
\usepackage{xspace}
\usepackage{xcolor}
\usepackage{soul}
\usepackage{natbib}
\usepackage{txfonts}
\usepackage{adjustbox}
\usepackage[version=4]{mhchem}
\bibpunct{(}{)}{;}{a}{}{,} 

\usepackage{txfonts,textcomp}
\usepackage[colorlinks=true,allcolors=blue]{hyperref}
\usepackage{etoolbox}
\makeatletter 
  \patchcmd{\NAT@citex}
    {\@citea\NAT@hyper@{%
      \NAT@nmfmt{\NAT@nm}%
      \hyper@natlinkbreak{\NAT@aysep\NAT@spacechar}{\@citeb\@extra@b@citeb}%
      \NAT@date}}
    {\@citea\NAT@nmfmt{\NAT@nm}%
    \NAT@aysep\NAT@spacechar\NAT@hyper@{\NAT@date}}{}{}

  \patchcmd{\NAT@citex}
    {\@citea\NAT@hyper@{%
      \NAT@nmfmt{\NAT@nm}%
      \hyper@natlinkbreak{\NAT@spacechar\NAT@@open\if*#1*\else#1\NAT@spacechar\fi}%
        {\@citeb\@extra@b@citeb}%
      \NAT@date}}
    {\@citea\NAT@nmfmt{\NAT@nm}%
    \NAT@spacechar\NAT@@open\if*#1*\else#1\NAT@spacechar\fi\NAT@hyper@{\NAT@date}}
    {}{}
\makeatother

\newcommand\Msun{\text{M}_{\astrosun}} 
 
\newcommand\Zsun{\text{Z}_{\astrosun}}

\newcommand\HI{\ion{H}{I}\xspace} 
\newcommand\HII{\ion{H}{II}\xspace} 
\newcommand\HeI{\ion{He}{I}\xspace} 
\newcommand\HeII{\ion{He}{II}\xspace}

\newcommand\arepo{\textsc{arepo}\xspace}
\newcommand\areport{\mbox{\textsc{arepo-rt}}\xspace}

\newcommand\orcid[1]{\protect\href{http://orcid.org/#1}{\protect\includegraphics[height=12pt]{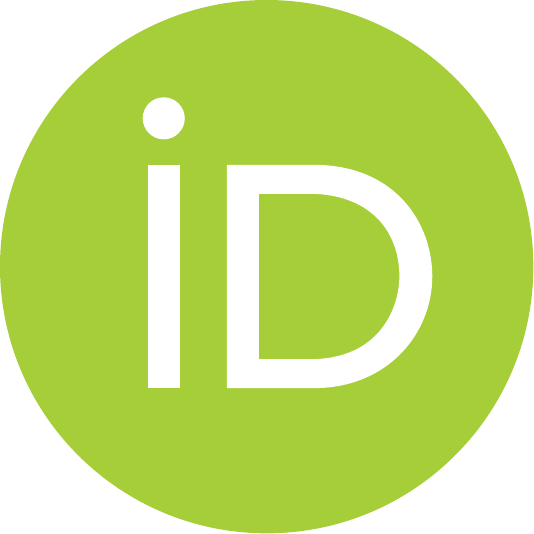}}}

\begin{document} 

   \title{RIGEL: Feedback-regulated cloud-scale star formation efficiency in a simulated dwarf galaxy merger}
   \authorrunning{Y. Deng et al.}
   \titlerunning{RIGEL: SFE in a dwarf starburst}

  \author{Yunwei Deng\inst{1}\orcid{0000-0002-7478-6427}
          \and Hui Li\inst{1}\orcid{0000-0002-1253-2763}
          \and Federico Marinacci\inst{2,3}\orcid{0000-0003-3816-7028}
           \and Yang Ni\inst{4}\orcid{0000-0003-0794-1949}
          \and Boyuan~Liu\inst{5}\orcid{0000-0002-4966-7450}
          \and Aaron~Smith\inst{6}\orcid{0000-0002-2838-9033}\and\\
          Rahul~Kannan\inst{7}\orcid{0000-0001-6092-2187}
          \and Greg~L.~Bryan\inst{8}\orcid{0000-0003-2630-9228}
          }

   \institute{
             1. Department of Astronomy, Tsinghua University, Haidian DS 100084, Beijing, China\\
             \email{\href{mailto:hliastro@tsinghua.edu.cn}{hliastro@tsinghua.edu.cn}}\\
             2. Department of Physics \& Astronomy ``Augusto Righi'', University of Bologna, via Gobetti 93/2, I-40129 Bologna, Italy\\
             3. INAF, Astrophysics and Space Science Observatory Bologna, Via P. Gobetti 93/3, I-40129 Bologna, Italy\\
             4. Institute for Advanced Study, Tsinghua University, Beijing 100084, China\\
             5. Institut f\"ur Theoretische Astrophysik, Zentrum f\"ur Astronomie, Universit\"at Heidelberg, D-69120 Heidelberg, Germany\\
             6. Department of Physics, The University of Texas at Dallas, Richardson, Texas 75080, USA\\
             7. Department of Physics and Astronomy, York University, 4700 Keele Street, Toronto, ON M3J 1P3, Canada\\
             8. Department of Astronomy, Columbia University, New York, NY 10027, USA
             }

   \date{Received XXX; accepted XXX}

\abstract
{Major mergers of galaxies are likely to trigger bursty star formation activities. Usually, the accumulation of dense gas and the boost of star formation efficiency (SFE) are considered to be the two main drivers of starbursts. However, it remains unclear how each process operates on the scale of individual star-forming clouds. Here, we present a high-resolution ($2\,\Msun$) radiation-hydrodynamic simulation of a gas-rich dwarf galaxy merger using the Realistic ISM modeling in Galaxy Evolution and Lifecycles (RIGEL) model to investigate how mergers affect the properties of the structure of dense star-forming gas and the cloud-scale SFE. With the unprecedented mass and temporal resolution of the simulations, we tracked the evolution of sub-virial dense clouds in the simulation by mapping them across successive snapshots spanning 200\,Myr taken at intervals of 0.2\,Myr. We find that the merger triggers a $130$ fold increase in the star formation rate (SFR) and shortens the galaxy-wide gas-depletion time by two orders of magnitude compared to those in two matched isolated galaxies. However, the depletion time of individual clouds and their lifetime distribution remained unchanged over the simulation period. The cloud life cycles and cloud-scale SFE are determined by local stellar feedback rather than such environmental factors as tidal fields regardless of the merger process, and the integrated SFE ($\epsilon_\text{int}$) of clouds in complex environments remains well-described by an $\epsilon_\text{int}$--$\Sigma_\text{tot}$ relation found in idealized isolated-cloud experiments.
During the peak of the starburst, the median cloud-scale integrated SFE was lower by only 0.17--0.33\,dex compared to the value when the two galaxies were not interacting. 
The merger boosts the SFR primarily through the accumulation and compression of dense gas fueling star formation. Strong tidal torques assemble $\gtrsim10^{5}\,\Msun$ clouds, which seed massive stellar clusters. The average separation between star-forming clouds decreases during the merger, which in turn decreases the cloud--cluster spatial de-correlation from $\gtrsim1$\,kpc to $\sim0.1$\,kpc depicted in tuning fork diagrams -- a testable prediction for future observations of interacting low-mass galaxies.}


   \maketitle

\defcitealias{2020MNRAS.495.4170T}{T20}
\defcitealias{2003ApJS..146..417L}{LH03}
\defcitealias{Schaerer02}{S02}
\defcitealias{Gessey-Jones2022}{GJ22}
\defcitealias{2019MNRAS.482.1304E}{E19}
\defcitealias{2017MNRAS.471.2151H}{H17}
\defcitealias{2023MNRAS.522.3092L}{L23}
\defcitealias{2018ApJS..237...13L}{LC18}
\defcitealias{2024A&A...691A.231D:Deng}{D24}
\defcitealias{2020ApJ...891....2L}{L20}
\section{Introduction}
Star formation (SF) in the observable universe is notably an inefficient process by any measure. On a galactic scale, only about $2\%$ of potential star-forming gas is actually converted into stars during each dynamical time \citep{1998ApJ...498..541K,2012ARA&A..50..531K:Kennicutt}. This low galaxy-wide star formation efficiency (SFE) is generally linked to feedback mechanisms from massive stars as well as active galactic nuclei (AGNs). Specifically, stellar feedback shapes the evolution and life cycles of star-forming regions at the scale of molecular clouds (MCs) and results in a very short period of SF for each MC of several megayears \citep{2015MNRAS.449.1106H,2019Natur.569..519K,2022MNRAS.509..272C,2023ASPC..534....1C}. Consequently, this regulates the cloud-scale SFE at a low level in regular star-forming environments \citep{2009ApJS..181..321E:Evans,2011ApJ...729..133M:Murray,2016ApJ...833..229L:Lee,2024A&A...688A.163M:Mattern,2025ApJ...985...14L:Leroy}.

However, systems with significantly enhanced galaxy-wide SFE are ubiquitous throughout the universe and are typically known as “starbursts” \citep{1981ApJ...248..105W:Weedman}. Major mergers of galaxies are commonly recognized as a triggering mechanism for starbursts based on observations \citep[e.g.,][]{1987AJ.....93.1011K:Kennicutt,1996ARA&A..34..749S:Sanders,2000ApJ...530..660B:Barton,2008AJ....135.1877E:Ellison,2021MNRAS.501..137H:Horstman,2023A&A...677A.179G:Gao,2025MNRAS.540..594K:Kaviraj} and numerical simulations (\citealt{1991ApJ...370L..65B:Barnes,2007A&A...468...61D:DiMatteo}; \citealt{2020ApJ...891....2L}, hereafter \citetalias{2020ApJ...891....2L}; \citealt{2019A&A...625A..65R:Renaud,2022MNRAS.516.4922R:Renaud}). The merger-driven starburst is an important stage for galaxy growth across cosmic time as well as the formation and evolution of globular cluster populations \citep{2017ApJ...834...69L:Li,2017MNRAS.465.3622R,2019ApJ...879L..18L,2024MNRAS.530..645L,2020MNRAS.495.4248K:Keller,2025A&A...699A..31P:Pascale}.

Observationally, merger-driven bursts are evident from the decreased depletion time $\tau_{\rm dep}$ (i.e., the inverse of the galaxy-wide SFE; see Sec.~\ref{sec:tdep_sfe}) on the Kennicutt--Schmidt plane and an enhanced amount of molecular gas traced by CO on scales ranging from the entire merger systems down to kiloparsec \citep[e.g.,][]{2010ApJ...714L.118D:Daddi,2010A&A...518L..44K:Klaas,2012A&A...540A..96M:Martinez-Badenes,2018ApJ...868..132P:Pan,2019AJ....157..131B:Bemis,2019ApJ...882....5W:Wilson,2021ApJ...908...61K:Kennicutt}. Recently, observations performed with the Atacama Large Millimeter/submillimeter Array (ALMA) have been able to resolve the molecular gas in nearby merger and starburst systems at a spatial resolution of $<100$\,pc \citep[e.g.,][]{2021MNRAS.500.4730B:Brunetti,2022MNRAS.515.2928B:Brunetti,2024MNRAS.530..597B:Brunetti}. The MCs in such galaxies as NGC~3256 and NGC~4038/9 are found to be large, dense, and highly turbulent. However, at the scale of individual MCs and star-forming regions, it remains unclear whether the galaxy-wide starburst in highly turbulent, out-of-equilibrium mergers is primarily driven by an enhanced cloud-scale SFE, an increased gas supply, or both.

Previous numerical studies have demonstrated that intense tidal forces during galaxy mergers enhance the formation of massive star clusters \citep[e.g.,][]{2008MNRAS.391L..98R:Renaud,2022MNRAS.514..265L:Li} while also quickly leading to their disruption \citep[e.g.,][]{2003MNRAS.340..227B:Baumgardt,2011MNRAS.414.1339K:Kruijssen}. The enhancement of massive cluster formation is attributed to tidal compression, which assembles cold and dense gas, whereas tidal disruption is a natural result of the stretching effect of the strong and variable external field. These dual effects are anticipated to influence the life cycle and ultimately the integrated SFE of the MCs within merger systems. While it is well-established that the merger-driven enhancement of dense gas fuels SF, grasping how these positive and negative effects of tides influence individual clouds is vital for physically comprehending merger-induced starburst events at the smallest scale. However, the complex hydrodynamical evolution of clouds and their fragility to stellar feedback render analytical modeling impractical.

In light of recent progress in observing and simulating the detailed multiphase interstellar medium (ISM) and SF within various galaxies, we present an idealized simulation of a merger between dwarf galaxies. This simulation used the cutting-edge galaxy formation model, Realistic ISM modeling in Galaxy Evolution and Lifecycles (RIGEL, \citealt{2024A&A...691A.231D:Deng}, henceforth referred to as \citetalias{2024A&A...691A.231D:Deng}), to present a numerical view of how dense clouds evolve and how stars form in a merger-induced starburst system with both high mass and temporal resolutions of $2\,\Msun$ and $0.2\,$Myr, respectively. The aim of this paper is to quantify the driving mechanisms of merger-induced starbursts and study both the properties of dense cloud populations and the relationship between cloud-scale SFE in different galactic environments.

The remainder of the paper is organized as follows. In Section~\ref{sec:method} we briefly summarize the physical processes in the RIGEL model, describe the initial conditions (ICs) of the dwarf--dwarf merger simulation, and illustrate the workflow to identify dense clouds and star clusters from the simulation snapshots. In Section~\ref{sec:overview} we give an overview of the global properties of the simulated galaxies. In Section~\ref{sec:cloud-scale} we zoom in on the cloud scale to study the properties of star-forming clouds and their integrated SFE. In Section~\ref{sec:tuningfork} we use the tuning fork diagrams to provide sub-kiloparsec scale observational predictions of the ISM structures in merger systems. Finally, we discuss the implications and caveats of our model and summarize the key results of the paper in Section~\ref{sec:discuss}.

\section{Method}
\label{sec:method}
The simulations in this work are performed with \arepo \citep{2010MNRAS.401..791S,2016MNRAS.455.1134P,2020ApJS..248...32W:Weinberger}, a moving-mesh, finite-volume hydrodynamic code with a second-order Godunov scheme. Following \citetalias{2024A&A...691A.231D:Deng}, the idealized galaxy merger simulation used the RIGEL framework, which includes self-gravity, hydrodynamics, radiative transfer, heating, cooling, SF, and stellar feedback through radiation, stellar winds, and supernovae (SNe) explosion. The initial condition for the merger is set the same as the one in \citetalias{2020ApJ...891....2L} for a direct comparison. We describe the RIGEL model and the ICs in detail in the following subsections.

\subsection{The RIGEL model}
\label{sec:RIGEL}

RIGEL tracks the formation and evolution of individual massive stars drawn from the \cite{2003PASP..115..763C} initial mass function (IMF) in the resolved multiphase ISM. Stars are formed in cold ($T <
T_\text{th}$), dense ($n_\text{H} > n_\text{th}$), contracting ($\nabla\cdot {\bm v} < 0$), self-gravitating, and marginally Jeans-resolved gas where the thermal Jeans length is smaller than four times the cell size ($L_\text{J}=\sqrt{\pi c_s^2 / G \rho}<4\Delta x$). Compared to \citetalias{2024A&A...691A.231D:Deng}, we used a new criterion to find the locally self-gravitating gas cells by $[||\nabla {\bm v}_i||^2+(c_{s,i}/\Delta x)^2)]/8\pi G\rho_i<f_\alpha$ \citep{2019MNRAS.489.4233M}, where $||\nabla {\bm v}_i||$ is the Frobenius norm of velocity and $f_\alpha$ is a free parameter. We adopted $f_\alpha=0.5$ through numerical experiments and found it properly selected the self-gravitating cells in gas above the threshold density. In this work given the mass resolution of $2\,\Msun$, the density and temperature thresholds for SF were set as $T_\text{th}=100$\,K and $n_\text{th}=3000$\,cm$^{-3}$, respectively. 

Lifetimes, photon production rates, mass-loss rates, and wind velocities of massive stars (stellar mass $M_\star > 8\,\Msun$) are determined by their initial masses and metallicities based on a library that incorporates a variety of stellar models. This stellar feedback model combines the stellar models covering the metallicity from $10^{-8}\,\Zsun$ to solar from \cite{Schaerer02}, \cite{2003ApJS..146..417L}, \cite{2019MNRAS.482.1304E}, \cite{2020MNRAS.495.4170T}, and \cite{Gessey-Jones2022}, and the details are described in Section~2.4 of \citetalias{2024A&A...691A.231D:Deng}. In our $2\,\Msun$ resolution simulation, the massive star mass drawn from IMF exceeds the mass of its host star particle. To ensure mass conservation at the star cluster scale, star particles continuously lose mass based on an IMF-averaged rate while still being able to release mass in discrete events such as supernovae, with any inconsistency from large ejecta masses being counterbalanced by the collective mass loss from all other star particles.

The detailed cooling in star-forming ISM \citep{2023ApJS..264...10K}, ionization feedback from \HII regions \citep{2024MNRAS.527..478D}, and the Sedov–Taylor phase of SNe \citep{2015ApJ...802...99K} were properly accounted for and resolved at a solar-mass-level resolution. The radiation field of seven spectral bands, including infrared (IR, $0.1-1$\,eV), optical (Opt., $1-5.8$\,eV), far-ultraviolet (FUV, $5.8-11.2$\,eV), Lyman–Warner (LW, $11.2-13.6$\,eV), hydrogen ionizing (EUV1, $13.6-24.6$\,eV), \HeI ionizing (EUV2, $24.6-54.4$\,eV), and \HeII ionizing (EUV3, $54.4-\infty$\,eV) bands \citep{2020MNRAS.499.5732K}, was explicitly modeled and coupled to gas hydrodynamics via the moment-based RHD solver \areport \citep{2019MNRAS.485..117K,2024MNRAS.533..268Z:Zier}.

\subsection{Initial condition}
\label{sec:IC}
We created a merger IC of two identical disks with the parameters identical to the IC used in \citetalias{2020ApJ...891....2L}. Each dwarf galaxy consists of $2\times10^{10}\,\Msun$ dark matter (DM) particles,
a $4\times10^7\,\Msun$ gas disk with a scale length of $1.46$\,kpc, and a $2\times10^7\,\Msun$ background stellar disk with a scale length of $0.73$\,kpc and a scale
height of $0.35$\,kpc. The DM halo of each galaxy follows a Hernquist profile with an Navarro–Frenk–White (NFW)-equivalent concentration parameter $c=10$ and a spin parameter $\lambda = 0.03$. The mass resolution was defined to $2500\,\Msun$ for DM and $2\,\Msun$ for baryonic components. The gravitational softening lengths were set to $\epsilon_\text{DM}=39$\,pc
for DM and $\epsilon_\star=0.05$\,pc for star particles, while the softening of gas particles was determined adaptively with a minimum length of 0.01\,pc. The two dwarf galaxies were set on parabolic orbits with a pericentric distance of $d_\text{peri}=1.46$\,kpc, an initial separation
of $d_\text{init}=5$\,kpc, and the
disk spin parameters of $(\theta_1,\phi_1,\theta_2,\phi_2)= (60^{\circ},30^{\circ},60^{\circ},240^{\circ})$. The initial metallicity of these two galaxies is $0.1\,\Zsun$. 

We ran the simulation for $200$\,Myr of the simulation time. To accurately link the clouds between consecutive snapshots and build the cloud evolution graph (see Section~\ref{sec:network}), we output snapshots every 0.2\,Myr.

\subsection{Cloud and cluster identification}
\label{sec:identify}
We used the {\sc CloudPhinder}\footnote[1]{\url{https://github.com/mikegrudic/CloudPhinder}} package to identify the gas clouds in our simulations. {\sc CloudPhinder} detects the local density peaks in galaxies, then identifies the largest self-gravitating gaseous structures by searching for neighboring cells. 

To identify the potential and active star-forming clouds, we only considered dense gas, i.e., gas cells with a density of $n_\text{H}>100$\,cm$^{-3}$, similar to the density threshold in previous work \citep[e.g.,][]{2018MNRAS.479.3167G:Grisdale,2025A&A...699A.282N:Ni,2024MNRAS.534..215F:Fotopoulou}. Furthermore, the virial parameter $\alpha$ for these detected gas structures had to be less than 10. This set of selection criteria was relatively lenient, as our aim was to capture the evolution of clouds throughout their entire life cycle. Similar to \cite{2024MNRAS.534..215F:Fotopoulou}, our metal-poor dwarf galaxies are deficient in \ce{H2} gas, and the large-scale reservoir for SF is cold atomic gas. Thus, we did not use \ce{H2} fraction to select star-forming clouds. 

To identify young star clusters, we used the built-in {\sc subfind} algorithm \citep{2001MNRAS.328..726S:Springel} of \arepo to identify bound stellar systems. Bound stellar systems with at least 35 member stars were identified as star clusters. Clusters with a median age of member stars $t_\text{50}<5$\,Myr were defined as young star clusters.

\subsection{Cloud and cluster evolution graph (merger tree) construction}
\label{sec:network}
After identifying the clouds and clusters in each simulation snapshot, we tracked their evolution as a function of time by linking their progenitors and descendants across successive snapshots. This procedure resembles constructing merger trees for halos in cosmological simulations; however, unlike halos, clouds and clusters frequently split and dissolve due to feedback and dynamical processes. In mathematics, such graphs are called directed acyclic graphs (DAGs). Here we refer to these graphs as an “evolution graph."

Each cloud can spawn multiple children (cloud splits) or have multiple parents (cloud mergers). Following the method outlined in \cite{2025A&A...699A.282N:Ni}, we connected the clouds that had more than one particle ID in common and stored the parents and children of every cloud using the {\sc Python} package {\sc networkx} \citep{Hagberg2008ExploringNS}. To prevent marginal connections, the parent and child cloud must have more than 1\% of their mass in common.

To generate the evolution paths of the clouds within the graph, we began to walk through the graph from the clouds without any predecessors, referred to as “root nodes.” 
These root nodes can have one or more child clouds.
Thus, each root node generated several possible evolution paths based on the options chosen from the multi-child nodes. In general, we used the mass fraction of inherited gas in each child cloud to describe the probability that the child cloud $k$ would become the next node in the path was $P_k = m^\text{inh}_k/\sum_i m^\text{inh}_i$, where $m^\text{inh}_k$ is the inherited mass cloud $k$ and $\sum_i m^\text{inh}_i$ is the total mass inherited by all children. 
We employed various techniques to examine the evolution graphs for particular queries. The specific methods are presented when addressing the relevant questions raised in this paper. The details of how to link clouds are described in Section~2.4 of \cite{2025A&A...699A.282N:Ni}.

\section{Global properties of the dwarf galaxy merger}
\label{sec:overview}

\begin{figure}
\includegraphics[width=1.0\columnwidth]{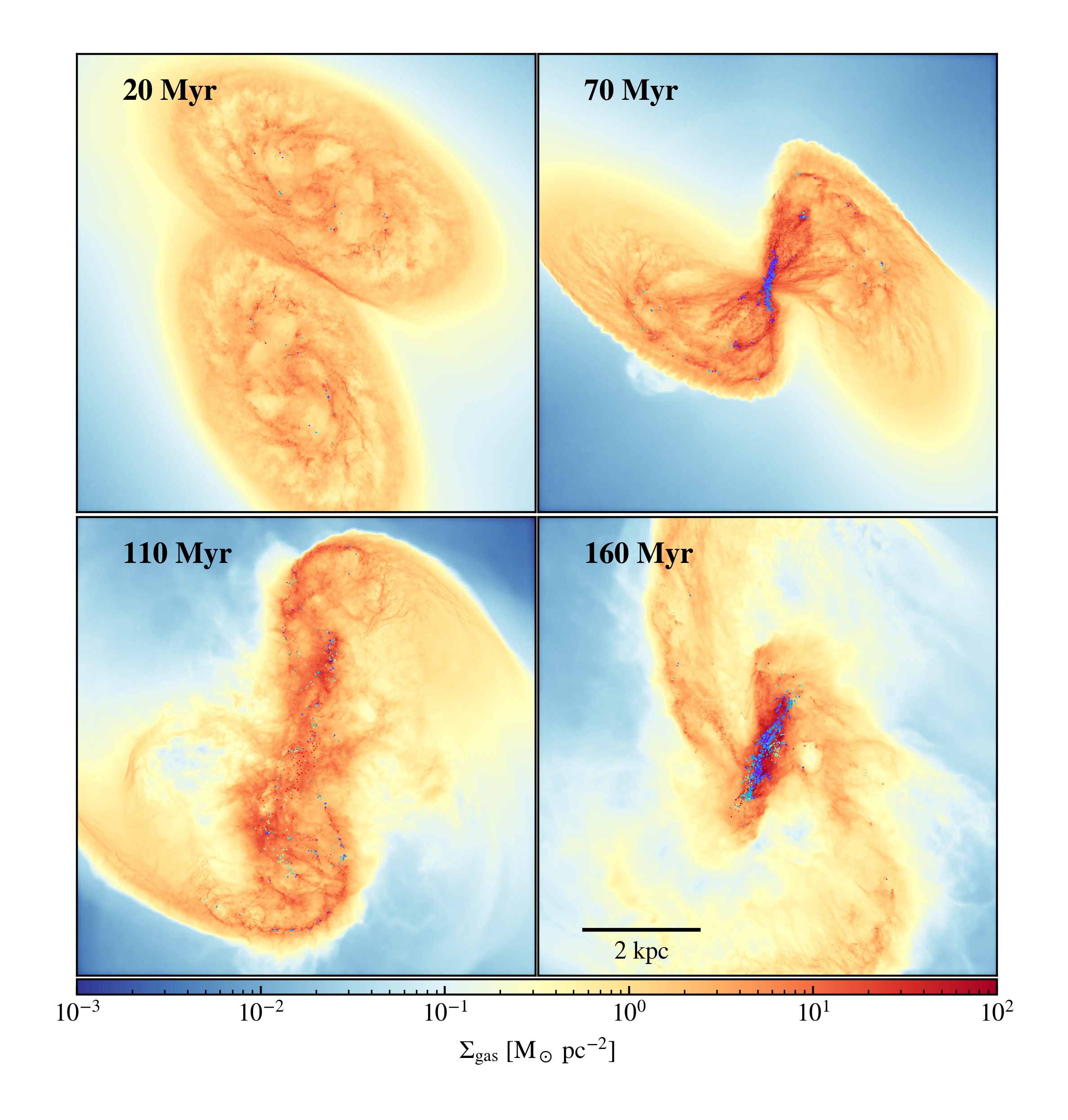}
    \caption{Gas surface density maps in different merger stages. The four panels show snapshots of the approaching stage (20\,Myr, upper left), the first passage when the first SF peak appears (70\,Myr, upper right), the first apocenter when the SFR exhibits a plateau (110\,Myr, bottom left), and the second encounter when the second SF peak occurs (160\,Myr, bottom right). The overset points represent young ($<40$\,Myr) and massive ($>8\,\Msun$) stars with ages color-coded from violet (youngest) to red (oldest). A movie of the simulation is available at \url{https://www.bilibili.com/video/BV18vKRzzEnj}.}
    \label{fig:projection}
\end{figure}

\begin{figure}
\includegraphics[width=\columnwidth]{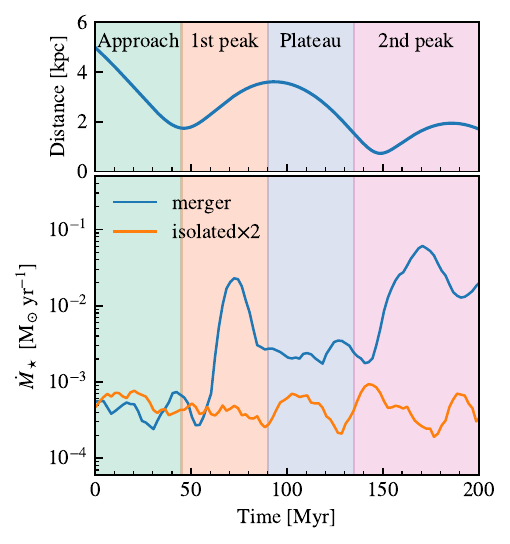}
    \caption{Evolution of the distance between two galaxies (upper panel) and the SF history of the merger system (bottom panel). The distance is measured between the centers of mass of the background stars in the galaxies. Bottom panel: SFR averaged over a timescale of 5\,Myr. The blue curve represents the total SFR of the two dwarfs in the merger, while the orange curve represents twice the SFR of an isolated dwarf galaxy. The four stages of the merger—“Approach,” “$1^\text{st}$ peak,” “Plateau,” and “$2^\text{nd}$ peak”—are marked with shaded regions in various colors.}
    \label{fig:SFR}
\end{figure}
\subsection{Global star formation rate}
In this section, we present an overview of the star formation activities in our simulation. Fig.~\ref{fig:projection} provides an overview of the merger process through the gas surface density distribution. Fig.~\ref{fig:SFR} shows the time evolution of the distance between the centers of the two galaxies (top panel) and the global star formation rate (SFR; bottom panel). The two galaxies have two pericentric passages during the 200\,Myr simulation time: the first at $\sim45$\,Myr, and the second at $\sim150$\,Myr. During the two close encounters, intense galactic torques and tidal forces induce a considerable inflow of gas toward the central 2\,kpc region of the merging system, resulting in significant SF in these gas-rich regions, as demonstrated by the emergence of young stars in the central interacting regions. As expected, the close encounters of galaxies significantly boost the SFR of the galaxies and induce two starbursts, onset at $\sim60$\,Myr and $\sim160$\,Myr, respectively. The SFR reaches $\gtrsim0.02\,\Msun$\,yr$^{-1 }$ during the first starburst and surges to a few $0.1\,\Msun$\,yr$^{-1}$ at the peak of the second starburst. The peak SFR during the starburst is 130 times the median SFR of two isolated dwarf galaxies. In our simulation, as the total gas mass decreases monotonically and is not significantly consumed by SF, a 130 times increase in the SFR directly correlates to a 130 times enhancement in the galaxy-wide SFE.

As the SFR varies across two orders of magnitudes, we divided the 200\,Myr of simulation time into four stages based on the SFR and merger progress to analyze the results. These phases are labeled as “Approach” for $[0,45)$\,Myr, “$1^\text{st}$ peak” for $[45,90)$\,Myr, “Plateau” for $[90,135)$\,Myr, and “$2^\text{nd}$ peak” for $[135,200)$\,Myr.

\subsection{Galaxy-wide depletion time and star formation efficiency}
\label{sec:tdep_sfe}
\begin{figure}
	\includegraphics[width=\columnwidth]{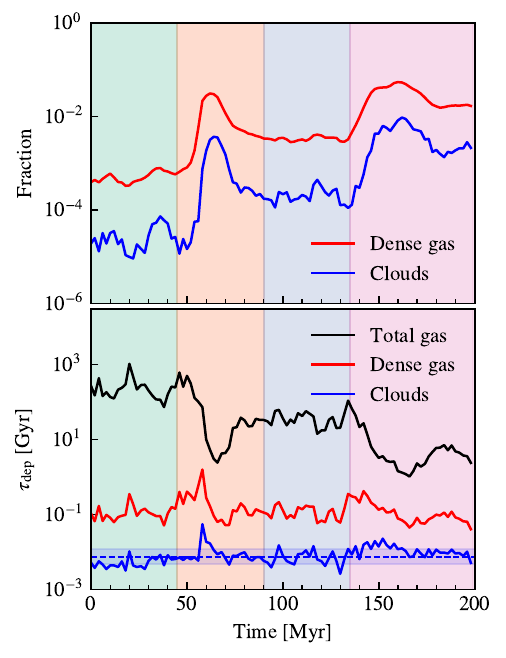}
    \caption{Evolution of galaxy-wide ISM properties and depletion time. Upper panel: Evolution of the fraction of $n_\text{H}>100$\,cm$^{-3}$ dense gas (red curve) and gas in identified clouds (dense gas clumps with virial parameter $\alpha<10$, blue curve). Bottom panel: Galaxy-wide depletion times calculated by the total gas mass (black curve), dense gas mass (red curve), and total cloud mass (blue curve). The depletion time of the gas in the clouds fluctuates around its median value of 7.4\,Myr (dashed blue line), with a 16--84 percentile range of 4.7--12\,Myr.
    }
    \label{fig:tdep}
\end{figure}

To study the triggering mechanism of the starburst in our dwarf galaxy merger, here we check the dense gas fraction in the galaxy and the galaxy-wide 
SFE.
In the upper panel of Fig.~\ref{fig:tdep}, we plot the fraction of $n_\text{H}>100$\,cm$^{-3}$ dense gas and gas in identified clouds. Similar to what \cite{2022MNRAS.514..265L:Li} found in the merger of Milky Way (MW)-like galaxies, the merger significantly increases the amount of dense gas and enhances the formation of clouds. During the second SF peak, the total mass of the dense gas is 56 times higher than that during the approach stage. The evolution of these two gas fractions highly resembles the evolution of the SFR, suggesting that the SFE of the dense gas is roughly constant.

Here, we use the galaxy-wide depletion time as an indicator of the galaxy-wide SFE. The depletion time $\tau_\text{dep} = M_\text{gas}/\dot{M}_\star$ is the inverse of the SFE per dynamical time. In Fig.~\ref{fig:tdep}, we present the depletion time for the total gas mass, dense gas mass, and total cloud mass, respectively. As expected, the total gas depletion time varies across two orders of magnitude from $\gtrsim100$\,Gyr to $\sim 1$\,Gyr. However, the depletion times of the dense gas and, especially, clouds do not exhibit as significant variation as that of the total gas. The depletion time for dense gas varies by up to a factor of 10. The depletion time of the gas in clouds fluctuates around its median value of 7.4\,Myr with a 16--84 percentile range of 4.7--12\,Myr. At the moment when the total gas depletion time significantly decreases (i.e., the $1^\text{st}$ and $2^\text{nd}$ SF peak), the cloud depletion time even increases and decays rapidly to the median value.

This short but universal dense gas/cloud depletion time indicates that on the galaxy scale, the higher SFE in the sense of total gas is solely because of the increased amount of dense gas, while the efficiency of dense gas converting to stars remains unchanged. However, it is expected that the strong tides and compression during the merger should impact the properties of dense gas and clouds. To understand this phenomenon, we zoomed-in on cloud scales to study how the merger affects the lifetimes and separations of clouds, as well as the cloud-scale integrated SFE.

\section{Zoom-in to cloud scale: Life cycle of clouds and integrated SFE}
\label{sec:cloud-scale}
\begin{figure*}
	\includegraphics[width=2\columnwidth]{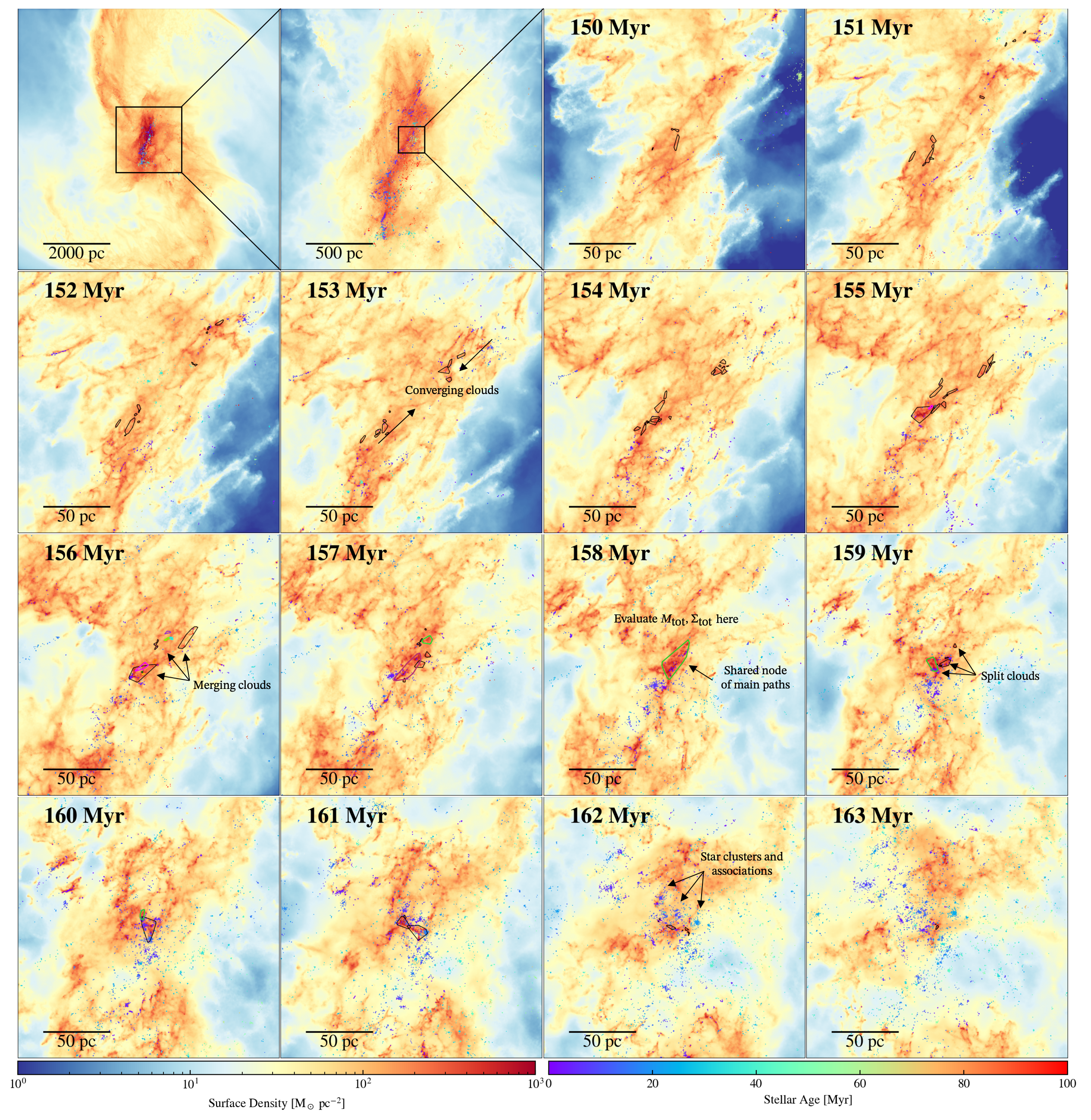}
    \caption{Evolution of clouds in a sequence spanning 14\,Myr. The first two panels show the location of the zoom-in region in the merger system. The third to the last panels show the column density distribution inside a (200\,pc)$^3$ box. In these panels, the color maps refer to the gas surface density distribution of the zoom-in regions, and the dots represent the stars color-coded according to their age. The polygons in the zoom-in panels depict the 2D convex hulls of the clouds identified from a specific evolution graph. The evolution paths in this graph merge at 158\,Myr and share a node (green polygon), where we evaluate the total baryon mass $M_\text{tot}$ and surface density $\Sigma_\text{tot}$ for the SFE calculation (see Section~\ref{sec:sfe}). The clouds form in the upper-right and lower-left regions at the beginning (see the panel of 153\,Myr for an example) and accumulate at the center (155\,Myr). Star clusters appear at 156\,Myr, but the cloud continues to accumulate mass from the upper-right, forming a large single cloud at 159\,Myr. The cloud continues to form stars while collapsing. Finally, the stellar feedback rapidly disperses the clouds, leaving bubbles within 4\,Myr. The colorful polygons appearing between 155\,Myr to 160\,Myr indicate the main paths sharing the same maximum baryon mass node in this evolution graph.
    }
    \label{fig:zoomin}
\end{figure*}
We use the evolution graph of the identified clouds to study the SF activities on the cloud scale. As an illustration, Fig.~\ref{fig:zoomin} shows the temporal evolution of the clouds in an evolution graph spanning 14\,Myr from 150\,Myr to 163\,Myr. This period corresponds to the sharp surge of SFR before it reaches the summit of the second peak. As seen in 150--153\,Myr (the $3^{\text{rd}}$ to $6^{\text{th}}$ panels), small, isolated clouds condense from the highly turbulent gas structure. These clouds keep accreting gas and move close to each other due to gravity. At 155\,Myr (the $8^{\text{th}}$ panel), a large cloud is assembled at the center and begins to spawn young stars. Star clusters appear at 156\,Myr (the $9^{\text{th}}$ panel), while the cloud continues to accumulate mass from the upper-right, forming a large single cloud at 159\,Myr (the $12^{\text{th}}$ panel). Another round of SF occurs in this large cloud. As SF and stellar feedback occur, this large cloud splits into several smaller clouds. Subsequently, stellar feedback rapidly disperses the clouds and creates a bubble within 4\,Myr.

This example demonstrates the high complexity of cloud evolution dynamics: it potentially encompasses numerous merger and split events, and exhibits multiple cycles of SF. These dynamic changes can only be captured owing to the high spatial and temporal resolutions provided by our simulation.

\subsection{The cloud population in the merger system}
\label{sec:cl-pop}
\begin{figure}
	\includegraphics[width=\columnwidth]{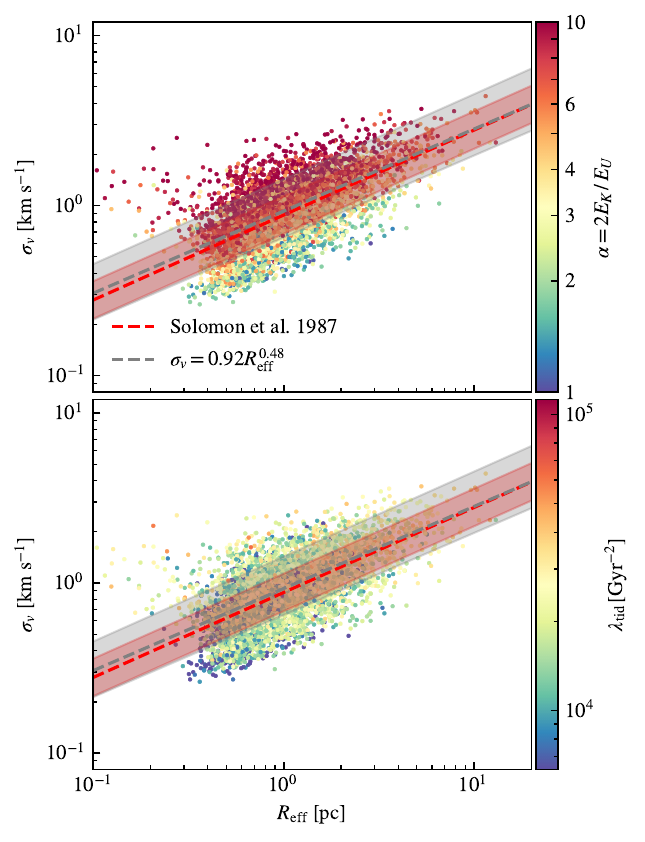}
    \caption{Cloud velocity dispersion as a function of effective radius. The clouds are selected from the snapshot spaced at 10\,Myr intervals to avoid duplicate counts and eliminate bias toward the long-lived clouds. The points in the top panel are color-coded by their virial parameters, while those in the bottom panel are color-coded by tidal strength $\lambda_\text{tid}$. Clouds in stronger tidal environments typically have larger velocity dispersions and virial parameters. The red line represents the observations of MW clouds by \cite{1987ApJ...319..730S:Solomon}, and the red-shaded region represents the observed 0.22\,dex dispersion of $\log_{10}\sigma_v$. The gray line and shaded region represent our best-fitting relation and the $\log_{10}\sigma_v$ dispersion of 0.31\,dex. Our sample fitting closely follows the observations, although the scatter in the sample is slightly larger than that of the observed MW clouds, due to a rather tolerant selection criterion designed to include dynamically diverse clouds.}
    \label{fig:Larson}
\end{figure}
We first give an overview of the physical properties of the clouds in our simulated merger system. In Fig.~\ref{fig:Larson}, we present the relation between the effective radius and velocity dispersion, a.k.a the Larson's “law” \citep{1981MNRAS.194..809L:Larson}. Following \cite{2025A&A...699A.282N:Ni}, we define the effective radius $R_\text{eff}$ and 1Dl velocity dispersion $\sigma_v$ of a cloud as
\begin{equation}
    R_\text{eff}=\sqrt{\frac{5}{3}\frac{\sum (m_i |\vec{x_i}-\vec{x_c}|^2)}{\sum m_i}},
\end{equation}
and
\begin{equation}
    \sigma_v=\sqrt{ \frac{\sum [m_i (|\vec{v_i}-\vec{v_c}|^2+c^2_{s,i})]}   {3\sum m_i} },
\end{equation}
where $\vec{x_c}$ and $\vec{v_c}$ are the center of mass (COM) position and velocity vector of the cloud, $m_i$, $c_{s,i}$, $\vec{v_i}$, and $\vec{x_i}$ are the mass, sound speed, velocity, and position vector of the $i$-th member gas cell of each cloud, respectively.

The cloud sample for Fig.~\ref{fig:Larson} is selected from simulation snapshots at different stages with a time interval of 10\,Myr, as it is long enough compared to the lifetime of most clouds in our simulations to avoid duplicate counts. Each point is color-coded by the virial parameter $\alpha = 2E_K/E_U$ (top panel) or the environmental tidal strength $\lambda_\text{tid}$ (bottom panel) estimated at a scale of 50\,pc following \cite{2019MNRAS.486.4030L:Li} and \cite{2022MNRAS.514..265L:Li}. The tidal strength $\lambda_\text{tid}$ is defined as the maximum of the absolute value of the eigenvalues of the tidal tensor $\bm{T}_{ij}$ \citep{2011MNRAS.418..759R:Renaud}. The tidal tensor of a cloud is defined at its COM ${\bm x}_\text{cl.}$ as
\begin{equation}
    \bm{T}_{ij} = \left.-\frac{\partial^2\Phi_\text{env.}(\bm{x})}{\partial x_i \partial x_j}\right|_{{\bm x} = {\bm x}_\text{cl.}}\,,
\end{equation}
where $\Phi_\text{env.}(\bm{x})=\Phi(\bm{x}) - \Phi_\text{cl.}(\bm{x})$ is the environmental gravitational potential, $\Phi(\bm{x})$ and $\Phi_\text{cl.}(\bm{x})$ are the total potential and the potential generated by the cloud itself, respectively. We used the {\sc Python} gravity solver {\sc pytreegrav} \citep{2021JOSS....6.3675G:Grudic} to evaluate the potentials.

The overall trend of the clouds and the best-fitting result of our sample closely follow the MW observations by \cite{1987ApJ...319..730S:Solomon}. However, the clouds exhibit a slightly larger scatter of 0.31\,dex on the Larson's relation compared to the observed scatter of 0.22\,dex. This is because we had a rather tolerant selection criterion, which allowed us to include dynamically different clouds into our sample. The virial parameter strongly correlates with the velocity dispersion as expected. The clouds with similar virial parameters roughly follow a similar relation with a similar slope. 

Moreover, we find that clouds with large velocity dispersion usually present in strong tidal fields, while those with small velocity dispersion are in weak tidal fields. For a given size, clouds in stronger tidal environments have higher velocity dispersions and are highly unbounded, suggesting that the clouds are more turbulent. Though the clouds are unbound as entities, the overdense substructures can collapse and finally form stars. This indicates a highly turbulent star-forming environment during the merger, akin to the observed cloud properties in merger systems \citep{2024MNRAS.530..597B:Brunetti}. Large clouds usually appear in stronger tidal environments, suggesting that the formation of larger clouds is facilitated by the strong tidal fields due to the merger. 

\begin{figure}
	\includegraphics[width=\columnwidth]{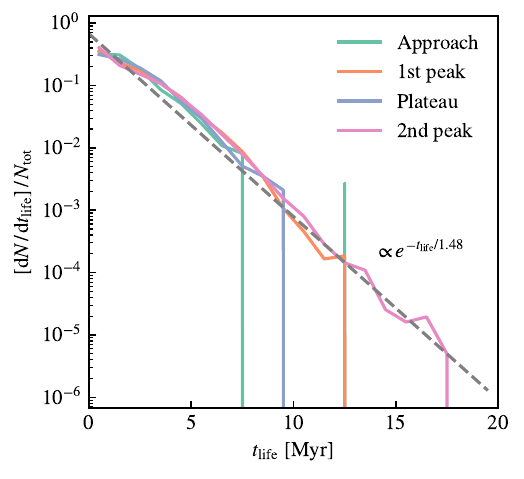}
    \caption{Cloud lifetime distributions at different merger stages. The dashed gray line denotes the maximum likelihood estimate fitting of the lifetime for all stages. The clouds in all merger stages follow the same exponentially decaying relation with a characteristic time of 1.48\,Myr.
    }
    \label{fig:lifetime}
\end{figure}

As the tidal force significantly affects the dynamics and boundness of the clouds \citep[e.g.,][]{2025ApJ...988..266L:Lee}, one may expect the lifetime of clouds in different merger stages to be different.  
To examine this, in Fig.~\ref{fig:lifetime}, we show the lifetime probability distribution functions (PDFs) of clouds in different merger stages. To statistically account for all identified clouds and their potential evolution paths, we used 300 Monte-Carlo walkers per root node to sample all possible evolution paths, resulting in 300 probability-weighted Monte-Carlo paths for each root node \citep[see Section 5.1 in][for details]{2021MNRAS.505.1678J:Jeffreson}. The lifetimes yielded by all the Monte-Carlo paths are considered as the sample for Fig.~\ref{fig:lifetime}.
Surprisingly, the normalized PDFs of the different stages follow the same exponentially decaying distribution. We fit the PDFs of all stages together and obtained a characteristic lifetime $\tau_\text{life} = 1.48$\,Myr.

To understand this, we first explain why the lifetime distribution is expected to be an exponentially decaying distribution. The exponential distributions arise in systems where the probability of a disruptive event occurring is constant over time, regardless of how long the system has existed. For the clouds, their dispersal can be attributed to various physical mechanisms including dynamical shearing, tidal disruption, SF, and stellar feedback (in the form of radiation, stellar winds, and supernovae). These mechanisms can be regarded as no memory and communication of each other in this context. The unified distribution of lifetimes across different stages suggests that the life cycles of our clouds barely depend on the galactic dynamics. On the contrary, the characteristic lifetime is dominated by the timescale of SF and feedback. Given the free-fall timescale at our cloud identification criterion $n_\text{H}=100$\,cm$^{-3}$ is 4\,Myr, it is unlikely to be determined by gas depletion due to SF. Therefore, the early stellar feedback through radiation and stellar winds is the key factor to determine the cloud lifetimes. This is also consistent with the observed feedback timescale of $1.5$\,Myr in NGC~300 \citep[see][]{2019Natur.569..519K}.

\subsection{Effects of compressive tidal fields on star-forming clouds}
\label{sec:compressive}

\begin{figure}
	\includegraphics[width=\columnwidth]{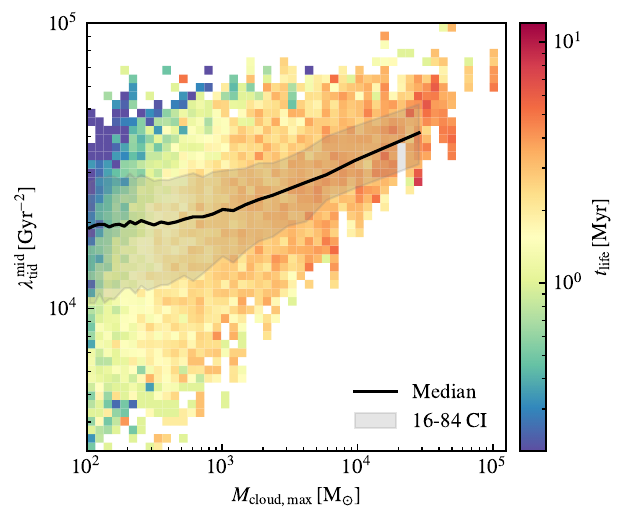     }
    \caption{Distribution of clouds in the $\lambda_\text{tid}^\text{mid}$-$M_\text{cloud,max}$ plane, color-coded by their average lifetimes. $\lambda_\text{tid}^\text{mid}$ is the median tidal strength experienced by clouds throughout their lifetime, and $M_\text{cloud,max}$ is the maximum mass the clouds ever reached during their lifetime. The solid black line and shaded region represent the median relation and the 16--84 percentile range, respectively. 
    }
    \label{fig:Mmax_tidal}
\end{figure}

As the tidal field is unlikely to dominate the lifetime of individual clouds, we now examine its effect on other cloud properties, such as their masses and average separations.
Figure~\ref{fig:Mmax_tidal} shows the distribution of the median tidal strength $\lambda_\text{tid}^\text{mid}$ and the maximum mass that clouds achieve throughout their lifetime, color-coded by the cloud lifetimes. We find that the maximum cloud mass correlates positively with the tidal strength, as the young cluster mass reported by \cite{2022MNRAS.514..265L:Li}. Massive clouds preferentially reside in strong tidal fields, while low-mass clouds have a spread distribution. This is because only the overdense regions in compressive tidal fields can keep accreting and assembling enough dense gas to collapse and form such massive bound structures.

The low-mass clouds that reside in strong tidal fields exhibit shorter lifetimes. However, for the $M_\text{cloud,max}>1000\,\Msun$ clouds, this correlation is much less significant. Massive clouds located in strong tides can still be as long-lived as those in weak tides. For the most massive clouds, their lifetimes are also the longest. 

This result is plausible because the clouds we selected are (quasi-)self-gravitationally bound systems, so that the more massive ones are less susceptible to the galactic tidal fields. Therefore, stellar feedback, which is much more rapid and powerful, still plays the key role in dispersing these clouds. Consequently, the cloud lifetimes exhibit a unified characteristic timescale of $1.48$\,Myr across all the merger stages. 

\begin{figure}
	\includegraphics[width=\columnwidth]{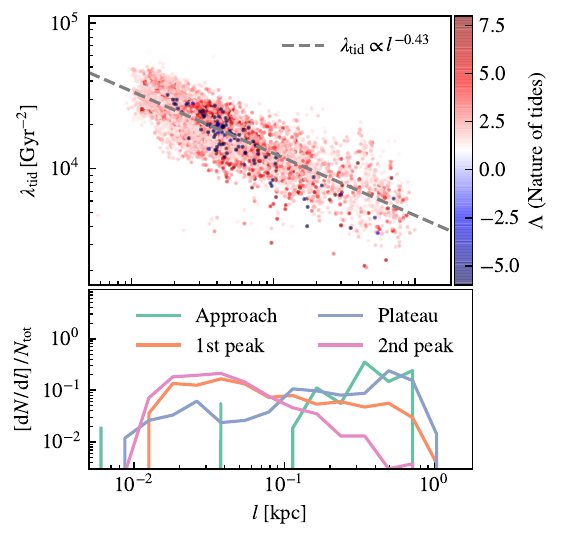}
    \caption{Relation between average cloud separation $l$ and tidal strength. The cloud sample is the same as that presented in Fig.~\ref{fig:Larson}. The top panel is color-coded by the nature of tides $\Lambda$. The dashed gray line represents the fitting of $\lambda_\text{tid}\propto l^{-0.43}$. The bottom panel shows the PDFs of the average separation $l$ in different merger stages. Most of the clouds are in fully compressive tidal fields ($\Lambda>1$). The average separation  $l$ extends and shifts to smaller scales as the merger progresses and the SFR increases.
    }
    \label{fig:separation}
\end{figure}

To quantify the separation among individual clouds, we defined the average separation $l$ of a cloud as the mean distance to its 16 nearest neighbors. In the top panel of Fig.~\ref{fig:separation}, we show the relation between the tidal strength $\lambda_\text{tid}$ and average separation $l$. The data points are color-coded by the nature of the tidal field (extensive or compressive) experienced by the clouds. The nature of the tidal field is characterized by $\Lambda =(\lambda_2+\lambda_3 )/2\lambda_1$, where $\lambda_{1,2,3}$ are the largest, middle, and smallest eigenvalues of the tidal tensor ${\bm T}_{ij}$. Basically, $\Lambda>1$ indicates fully compressive tides, while $\Lambda<0$ indicates extensive tides \citep[see][Section~3.3 for details]{2017MNRAS.465.3622R}. 

Apparently, the majority of clouds exist within completely compressive tidal fields. This is expected, as dense clouds predominantly form in central areas of gravitational potential where the gradient is relatively flat \citep{renaud2009fully}. We observed a significant inverse relationship between $\lambda_\text{tid}$ and $l$, which can be described by the relationship $\lambda_\text{tid}\propto l^{-0.43}$. This negative correlation indicates that intense tidal fields compress the gas accumulated and confine the effect of stellar feedback in the central regions, thereby shortening the distance between star-forming clouds.

The bottom panel of Fig.~\ref{fig:separation} shows the PDFs of average separation $l$ in different merger stages. During the approach stage, the average separation $l$ mainly distributes in the $0.1$--$1$\,kpc with a median of 288\,pc. With the merger in progress and SFR increasing, $l$ extends and shifts to smaller scales. The average separations in the first and second peaks are 46\,pc and 32\,pc, respectively. These results indicate an increase in volume density of individual clouds at the kiloparsec scale during the merger, which contributes to the reduced gas-star de-correlation found in Section~\ref{sec:tuningfork}.

From the results presented in Section~\ref{sec:cl-pop} and \ref{sec:compressive}, we learn that compressive tidal fields facilitate the formation of dense gas clouds by compressing and fueling gas in the central regions, so that we observed the reduced separation among clouds. However, once the self-gravitating clouds have formed, their lifetimes remain dominated by the feedback from the stars they nurture.

\subsection{Star formation efficiency}
\label{sec:sfe}

We now study the cloud-scale SFE for our cloud sample. We use the integrated SFE, $\epsilon_\text{int} = M_\star/M_0$, to characterize the efficiency of SF for each cloud, where $M_\star$ is the final mass of stars formed and $M_0$ is the mass of the initial gas cloud. 

Idealized simulations of isolated GMCs have revealed that the cloud-scale integrated SFE is positively related to the initial mass $M_0$ or surface density $\Sigma_0$ of the GMCs \citep[e.g.,][]{2010ApJ...710L.142F:Fall,2018MNRAS.475.3511G:Grudic,2019MNRAS.487..364L}. Analytically, this can be understood through the force balance between gravitational collapse and the feedback-driven momentum flux. Assuming a spherically symmetric geometry where $M_\star$ stars at the center push a gas shell with a mass of $M_\text{sh}$ through their feedback, the balance equation goes \citep{2019MNRAS.487..364L}:
\begin{equation}
    \frac{M_\star\dot{p}_\star}{4\pi R^2}= \frac{GM_\star\Sigma_\text{sh}}{R^2}+\frac{\beta GM_\text{sh}\Sigma_\text{sh}}{R^2}\,,
\end{equation}
where $\Sigma_\text{sh}$ is the surface density of the gas shell, $\dot{p}_\star$ is the specific feedback momentum flux and $\beta$ is a geometric factor. By defining $\Gamma = \pi G\Sigma_0/(4f_\text{boost}\dot{p}_\star)$, which describes the relative strength between gravity and feedback, we can solve $\epsilon_\text{int} = M_\star/M_0$ as
\begin{equation}
\label{equ:eps}
\epsilon_\text{int} = \frac{\sqrt{\Gamma^2+(4\beta-2)\Gamma+1}-(2\beta-1)\Gamma-1}{2(1-\beta)\Gamma}\,,
\end{equation}
where $f_\text{boost}$ is a free parameter reflecting the strength of feedback compared to the fiducial value. \cite{2019MNRAS.487..364L} provided the best-fit values for the free parameters of $\beta = 1.83\pm0.89$ and $\dot{p}_\star = (3.32\pm0.64)\times10^9$\,cm\,s$^{-2}$. This equation goes as $\epsilon_\text{int}\rightarrow1$ when $\Sigma_0\rightarrow\infty$, suggesting that stellar feedback will fail at very high gas surface density \citep{2018MNRAS.475.3511G:Grudic,2023MNRAS.523.3201D:Dekel}. However, these types of theoretical models have rarely been tested in a more realistic galactic environment such as our merger system. 
\label{sec:SFE}

\begin{figure*}
	\includegraphics[width=2\columnwidth]{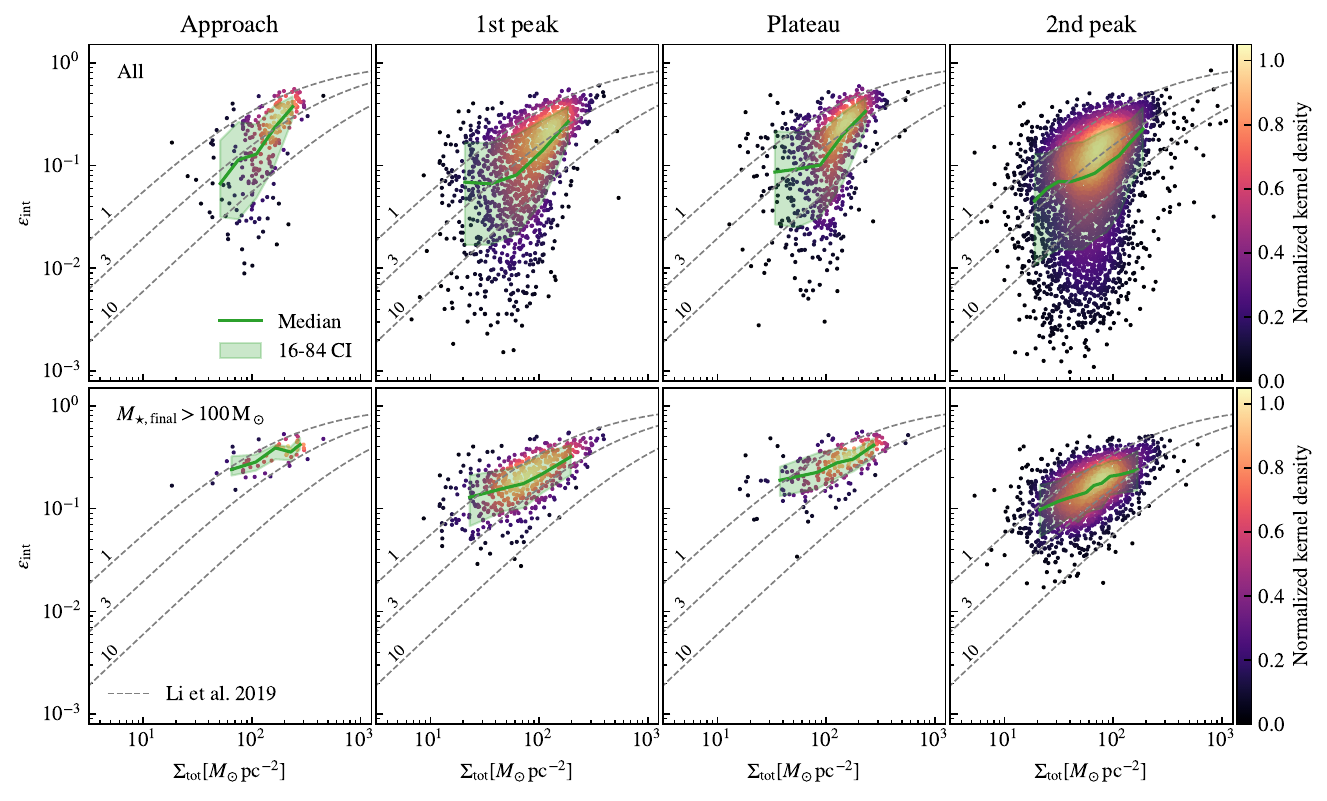}
    \caption{Distribution of integrated SFE $\epsilon_\text{int}$ as a function of the total baryon surface density  $\Sigma_\text{tot}$. Top panels: Distribution of all cloud samples with SF. Bottom panel: Results for the clouds formed more than $100\,\Msun$ in newly formed stars. The data points are color-coded by the normalized density in log-space estimated by a Gaussian kernel. The green curves show the median values, and the shaded regions indicate the 16--84 confidence levels. The dashed curves denote equation~(\ref{equ:eps}) with the best-fitting parameters from \cite{2019MNRAS.487..364L} for $f_\text{boost} = 1$, $3$, and $10$. Although the scatter of the simulated results is large, the median curve roughly follows the theoretical curve with $f_\text{boost} = 3$. The 16--84 confidence level lies between the $f_\text{boost} = 1$ and $10$ curves at all the stages.}
    \label{fig:SFE}
\end{figure*}

Here we use our high-resolution simulation to study the cloud-scale SFE. To do this, we must define the related quantities properly. As presented in Fig.~\ref{fig:zoomin}, the dynamics of cloud evolution are highly complex. Clouds linked in a single graph can undergo multiple times of merging and splitting, and even multiple cycles of SF. To obtain the integrated SFE, we first define the main evolution path of a cloud following \cite{2025A&A...699A.282N:Ni}. Given a root node, its main evolution path is defined as the evolution path with the maximum probability of 
\begin{equation}
    P(X_0 \rightarrow X_1 \rightarrow X_2 \rightarrow ... \rightarrow X_n)= \prod_{i=0}^{n-1} P(X_{i} \rightarrow X_{i+1}),
\end{equation}
where $P(X_{i} \rightarrow X_{i+1})$ is the probability of linking cloud $X_{i}$ and $X_{i+1}$.

The main evolution path prefers the shorter paths, so that it avoids including multiple rounds of SF. However, the main paths starting from different root nodes can merge together (i.e., small clouds assemble a large one), leading to duplicated counting.

To address this problem, we identify the node with the highest total baryon mass $M_\text{tot} = M_\text{gas}+M_\star$ for every main path, and we group the main paths that achieve this maximum baryon mass at the same node. For example, the polygons with different colors in Fig.~\ref{fig:zoomin} belong to different main paths. These main paths finally merge together to one large cloud at $158$\,Myr, and this cloud is thus the shared node of these main paths with the maximum baryon mass. The total baryon mass $M_\text{tot}$ and surface density for a cloud  $\Sigma_\text{tot}=M_\text{tot}/\pi R_\text{eff}^2$, are derived from the values of this shared node. Finally, the total new star mass $M_{\star,\text{final}}$ is calculated by summing the masses of new stars formed from nodes belonging to this group of main paths. The integrated SFE is $\epsilon_\text{int}=M_{\star,\text{final}}/M_\text{tot}$ as defined.

In Fig.~\ref{fig:SFE}, we present the distribution of the integrated SFE $\epsilon_\text{int}$ as a function of the total baryon surface density  $\Sigma_\text{tot}$. The distributions of all cloud samples with SF are shown in the top panels. We noticed that the theoretical models such as equation~(\ref{equ:eps}) require at least one massive star to provide feedback. Thus, we selected a subset of clouds, which form at least $100\,\Msun$ new stars, and the results are presented in the bottom panels of Fig.~\ref{fig:SFE}. We compare our results with the theoretical prediction given by equation~(\ref{equ:eps}) with the best-fitting parameters from \cite{2019MNRAS.487..364L} for $f_\text{boost} = 1$, $3$, and $10$. 

The highest integrated SFE within our simulated clouds is 84\%, while this cloud only forms $67\,\Msun$ stars. Thus, its high SFE is a natural result as it is free from stellar feedback. The lowest integrated SFE is 0.1\%, which also occurs in a low-mass cloud as a result of intense dynamical dispersal. Generally, the SFE of small clouds susceptible to dynamical effects strongly depends on the environments they reside in. 

For all the clouds, the scatter of the simulated results is large, the median curve roughly follows the theoretical curve with $f_\text{boost} = 3$, and the 16--84 confidence level lies between the $f_\text{boost} = 1$ and $10$ curves in all the stages. This large scatter is understandable as the SFE of numerous low-mass clouds is determined by environmental effects rather than stellar feedback.

On the contrary, once we confined our sample to the clouds with $M_{\star,\text{final}}>100\,\Msun$, the scatter becomes much smaller. By mass, 90\% of the stars are formed from such clouds. For these clouds, the highest integrated SFE is 60\% and the lowest is 1.7\%. The median curves and 16--84 confidence levels now lie between the $f_\text{boost} = 1$ and $3$ curves for all four merger stages. Quantitatively, within the range $30<\Sigma_\text{tot}/\Msun\,\text{pc}^{2}<300$, the median SFE during the second peak is 0.17--0.33\,dex (0.47 to 0.67 times) lower than that of the approach phase. This suggests that the strong tidal effects in mergers only slightly influence the cloud-scale SFE. This is consistent with what we find in cloud lifetimes. As a result, the $\epsilon_\text{int}$--$\Sigma_0$ relation found in the idealized cases can still be valid in complicated galactic environments. Since 90\% of the stars are formed in the clouds with $M_{\star,\text{final}}>100\,\Msun$, the majority of the SF activity is regulated by stellar feedback. This feedback-determined SFE reflects on the galaxy scale and results in the constant galaxy-wide cloud depletion time we found in Fig.~\ref{fig:tdep}.

\subsection{Tidal disruption of small clouds}
\label{sec:small-clouds}
\begin{figure}
	\includegraphics[width=\columnwidth]{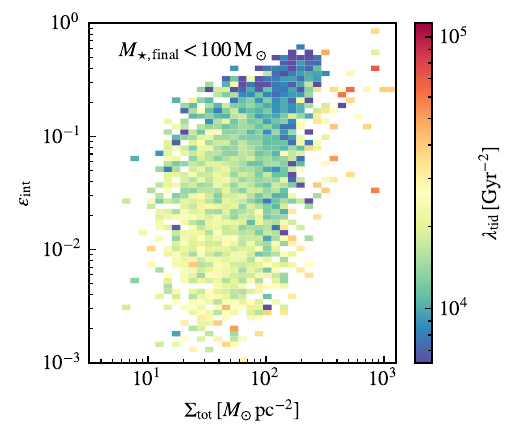     }
    \caption{Joint probability distribution of $\epsilon_\text{int}$ and $\Sigma_\text{tot}$ color-coded by the tidal strength. Here $\lambda_\text{tid}$ is the value when the cloud reaches its maximum baryon mass.
    }
    \label{fig:eps_tidal}
\end{figure}

We see in Fig.~\ref{fig:SFE} that the stellar feedback determines the integrated SFE of large clouds with $M_{\star,\text{final}}>100\,\Msun$. Although the small or low SFE clouds with $M_{\star,\text{final}}<100\,\Msun$ only form 10\% of the stars, it is interesting to examine whether the galactic dynamics regulates their SFE. 

In Fig.~\ref{fig:eps_tidal}, we present the joint probability distribution of $\epsilon_\text{int}$ and $\Sigma_\text{tot}$ color-coded by the tidal strength for the $M_{\star,\text{final}}<100\,\Msun$ small clouds. As these clouds are free from stellar feedback, their $\epsilon_\text{int}$ present a large scatter. Nonetheless, we can evidently see that for the $\Sigma_\text{tot}<300\,\Msun$\,pc$^{-2}$ clouds, the ones located in weak tidal fields present high integrated SFE reaching $70$\%, while the ones located in strong fields present low integrated SFE down to $0.1$\%. For these clouds, the strength of tidal disruption plays an essential role in determining their SFEs. The extensive distribution of $\epsilon_\text{int}$ reflects the drastic variation of tidal strength in the merger system.

Interestingly, several outliers with high surface density of $\Sigma_\text{tot}>300\,\Msun$\,pc$^{-2}$, high SFE of $\epsilon_\text{int}>10\%$, and high tidal strength of $\lambda_\text{tid}>3.5\times10^4$\,Gyr$^{-2}$ can be found in Fig.~\ref{fig:eps_tidal}. This may be due to the intense compressive tides causing these clouds to be sufficiently compact, allowing them to continue collapsing and forming stars without disturbance. We present an order-of-magnitude analysis about this in Section~\ref{sec:tidal-energy}.

\subsection{Formation of massive star clusters from massive clouds}
\begin{figure}
	\includegraphics[width=\columnwidth]{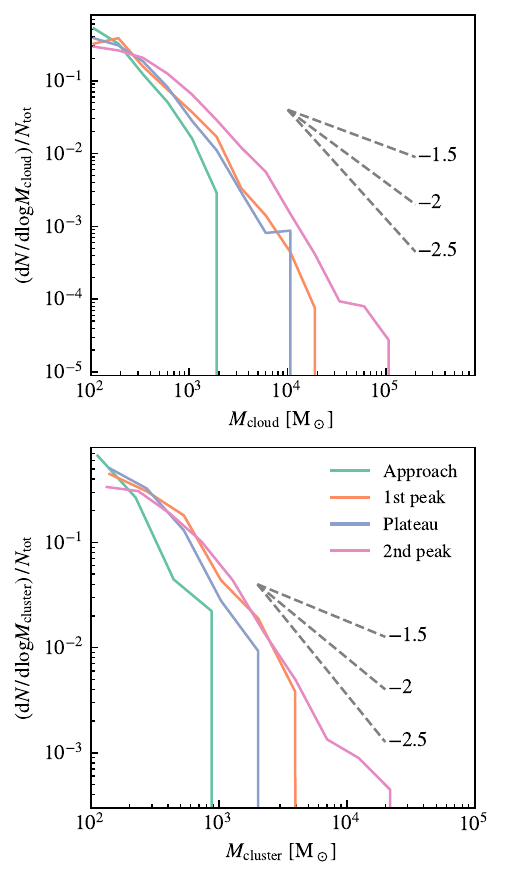}
    \caption{Cloud mass functions (top panel) and young cluster  ($<5$\,Myr) initial mass functions (CIMFs, bottom panel) of the merger system at different stages. 
    The dashed gray line denotes the power-law relation with a slope of $-1.5$, $-2$, and $-2.5$, respectively. A starburst induced by the merger leads to the formation of massive clusters, especially during the second SFR peak. The maximum cloud masses are $(1.8,\,23,\,11,\,90)\times10^3\,\Msun$ for the Approach, $1^\text{st}$ peak, Plateau, and $2^\text{nd}$ peak stages, respectively. The maximum cluster masses are $(0.93,\,3.4,\,1.8,\,26)\times10^3\,\Msun$ for these four stages.
    }
    \label{fig:MFs}
\end{figure}
Lastly, we examine the mass function of clouds and clusters as results of the varying galactic tidal fields. In Fig.~\ref{fig:MFs}, we present the cloud mass functions (top panel) and young cluster  ($<5$\,Myr) initial mass functions (CIMFs, bottom panel) of the merger system during different stages. 
To avoid duplicate counts, we selected clouds in snapshots with a time interval of 10\,Myr, which is much longer than the characteristic lifetime of our clouds (1.48\,Myr, see Section~\ref{sec:cl-pop}). On the other hand, the young clusters here are defined as the clusters whose median stellar age is younger than 5\,Myr, so the cadence to obtain CIMFs is 5\,Myr.

We describe the cloud mass functions and CIMFs as a power-law distribution. Evidently, the cloud mass functions during the merger-induced starburst exhibit a flatter slope and extend to a mass larger than $10^5\,\Msun$. Quantitatively, we fit the mass functions with a power-law form ${\rm d}N\,/\,{\rm d}M\propto M^{\alpha}$. For this fitting, we considered only clouds with masses exceeding $100\,\Msun$ to ensure a complete mass sampling. The power-law indices $\alpha$ and maximum cloud masses (captured by the selected snapshots) are $(-2.36,\,-1.84,\,-2.08,\,-1.79)$ and $(1.8,\,23,\,11,\,90)\times10^3\,\Msun$ for the Approach, $1^\text{st}$ peak, Plateau, and $2^\text{nd}$ peak stages, respectively. As the merger proceeds, dense gas accumulated in a stronger tidal field tends to form larger clouds (see Fig.~\ref{fig:Mmax_tidal}). Thus, the slope becomes shallower and the maximum cloud mass becomes larger. As a result, during the merger, there are more massive clouds that provide the fuel for the intense SF. Clouds dozens of times more massive than those in isolated galaxies appear during the merger and become the place to nurture massive star clusters.

As a result, the CIMF changes across stages with a similar trend, with the slopes of $(-2.23,\,-2.08,\,-2.32,\,-1.98)$ for each stage. This slope does not change as significantly as that of cloud mass functions, suggesting some self-regulation of the cluster formation. However, the maximum cluster mass does change significantly from $926\,\Msun$ within the approach stage to $2.6\times10^4\,\Msun$ within the $2^\text{nd}$ peak. We notice that the massive star clusters are still deficient compared with the increase in massive clouds, especially compared with the results of \citetalias{2020ApJ...891....2L}. We discuss this in Section~\ref{sec:comp}.

\begin{figure}
	\includegraphics[width=\columnwidth]{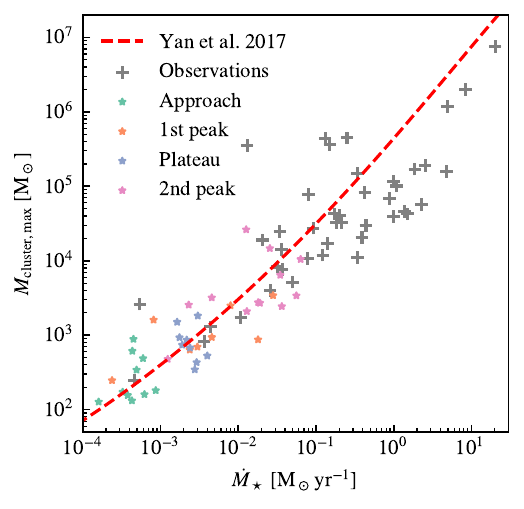}
    \caption{Most massive young cluster mass vs. galaxy-wide SFR ($M_\text{cluster,max}$–$\dot{M}_\star$) relation. The simulation data are color-coded by stage of the merger. The dashed red line represents the theoretical relation derived following \cite{2017A&A...607A.126Y:Yan}. The gray crosses represent observational data from \cite{2004MNRAS.348..187W:Weidner} and \cite{2015A&A...582A..93S:Schulz}, which have a typical uncertainty of 0.3 dex.
    }
    \label{fig:SFR-Mcl}
\end{figure}

An observation constraint of the cluster formation in different environments is the relation between the galaxy-wide SFR and maximum embedded cluster mass ($M_\text{cluster,max}$) in the galaxy \citep[e.g.,][]{2004MNRAS.348..187W:Weidner,2015A&A...582A..93S:Schulz}. This relation can be interpreted as a consequence of how the shape of the CIMF is influenced by SFR \citep{2013MNRAS.436.3309W:Weidner,2017ApJ...834...69L:Li}. We examine this relation in our simulation in Fig.~\ref{fig:SFR-Mcl}. Our simulated results roughly follow the theoretical relation derived by \cite{2017A&A...607A.126Y:Yan}, who assumed a single slope power-law embedded cluster mass function with an SFR-dependent power-law index \citep[e.g.,][]{2013MNRAS.436.3309W:Weidner}. Although our simulation only covered the low SFR range with a handful of data, it matches both the slope and the magnitude of the observations. Moreover, the most intense SFR during the $2^\text{nd}$ peak reaches the region with abundant observations, and the most massive young clusters formed in this period also present comparable masses with the observations. This result exhibits the ability and fidelity of the RIGEL model to capture and predict the self-regulated SF in drastically variable environments. We notice that the observations are mostly from more massive galaxies and not specifically selected to be mergers. A future application of RIGEL across a broader range of systems will allow for a more equitable comparison with observations.

\section{Spatial de-correlation between clouds and young clusters}
\label{sec:tuningfork}
In the previous section, we found that strong tides in mergers decrease the average separation between dense clouds but do not significantly change the efficiency of stellar feedback. This finding can provide observational predictions for future observations. 

In this section, we use the “tuning fork” diagram \citep{2018MNRAS.479.1866K:Kruijssen} to quantify the ISM structure and star-forming activities on scales of 0.1--1\,kpc, where the internal structures of clouds are not yet resolved. This diagram illustrates the relative bias of molecular gas depletion time measured in the apertures focused on either dense gas or young stars, making it an observational tool for measuring the spatial de-correlation between newly formed stellar regions and dense gas clouds. 

We noticed that the observational lower limit for molecular gas surface densities of $\Sigma_\text{\ce{H2}}>13\,\Msun$\,pc$^{-2}$ \citep{2019Natur.569..519K} is much higher than the typical $\Sigma_\text{\ce{H2}}$ in our simulation. Thus, we did not apply realistic observational sensitivity cuts to our analysis. The results obtained here can be regarded as a theoretical exploration and more practical observational predictions require future simulations of larger systems.

\begin{figure}
	\includegraphics[width=\columnwidth]{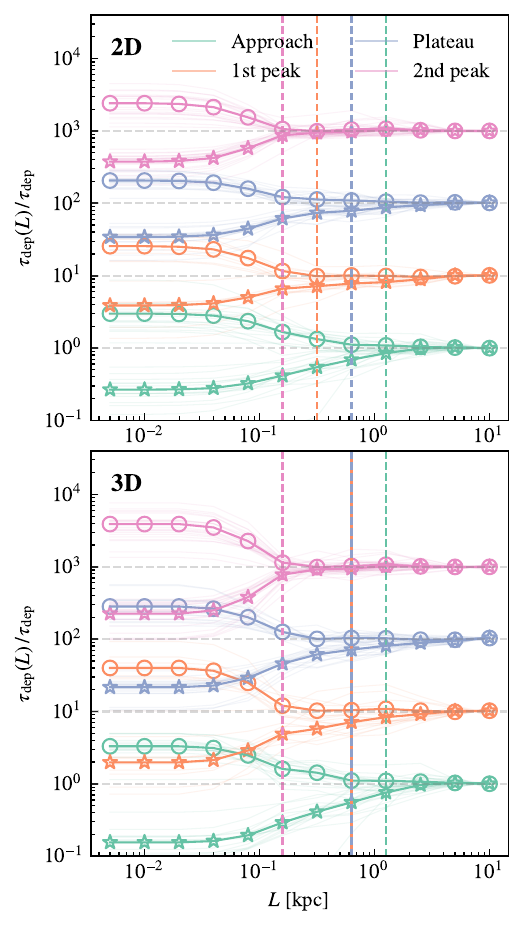}
    \caption{2D (top panel) and 3D (bottom panel) “tuning fork” diagrams from our simulation at different phases. These diagrams reflect the deviation of the \ce{H2} depletion time centered on \ce{H2} ($\tau_\text{\ce{H2},gas}$, upper branch) or recent SFR peaks ($\tau_\text{\ce{H2},SFR}$, lower branch) from the galaxy-averaged value as a function of the spatial scale $L$. The curves of the $1^\text{st}$ peak, Plateau, and $2^\text{nd}$ peak are multiplied by constants of $10$, $100$, and $1000$, respectively, to avoid overlapping. The vertical lines indicate the scale when $\tau_\text{\ce{H2},gas}$ exceeds $\tau_\text{\ce{H2},SFR}$ by 50\%.
    }
    \label{fig:tuingfork}
\end{figure}

\subsection{2D tuning fork diagram}
To generate the molecular gas map, we projected the volume density of \ce{H2} within an (8\,kpc)$^3$ box along the $z$-axis. In our $0.1\,\Zsun$ metal-poor galaxies, a significant fraction of the \ce{H2} mass can be found in diffuse, \HI-dominated gas where SF does not happen \citep[e.g.,][]{2021ApJ...920...44H,2024A&A...691A.231D:Deng}. We applied a cut on the \ce{H2} fraction of gas cells of $2x_\text{\ce{H2}}>0.1$ to exclude such diffuse \ce{H2}.

We followed the method outlined by \cite{2021ApJ...918...13S:Semenov} to generate the SFR map. We selected young star particles with ages 2--5\,Myr. We projected the young stars to the same mesh as the molecular gas map and compute $\dot{\Sigma}_\star$ in each pixel. The lower age cut mimics the typical timescale of the embedded phase of young clusters \citep[e.g.,][]{2021ApJ...911..128K,2024ApJ...974L..24D:Deshmukh}, while the upper limit mimics the typical stellar age traced by H$\alpha$ \citep{2021MNRAS.501.4812F:FloresVelazquez,SmithMW2022,2022MNRAS.513.2904T:Tacchella}.

With the $\Sigma_\text{\ce{H2}}$ and $\dot{\Sigma}_\star$ maps at a native resolution of 5\,pc, we smoothed the maps using a 2D Gaussian filter with a width of 20\,pc. We use the {\sc scikit-image} package \citep{van2014scikit} to identify the local extrema of $\Sigma_\text{\ce{H2}}$ and $\dot{\Sigma}_\star$ maps.
Following the method outlined by \cite{2018MNRAS.479.1866K:Kruijssen}, we then smoothed the maps with a top hat filter with the window size $L$ and computed the average $\Sigma_\text{\ce{H2}}$ and $\dot{\Sigma}_\star$ at the locations of gas and SFR peaks identified above. By doing this, we obtained a tuning fork diagram for a single snapshot.

We present the 2D tuning fork diagram from our simulation in the top panel of Fig.~\ref{fig:tuingfork}. In all four stages, our simulation results present the tuning-fork-like shape, while the scale at which the SFR branch and the gas branch converge differs with the stages. During the approach stage, the branches converge at the largest scale of $L\gtrsim1$\,kpc, which aligns with the observations of nearby disk galaxies \citep{2019Natur.569..519K,2022MNRAS.509..272C}. 

As the two galaxies merge, the convergence point shifts toward smaller scales. At the first SF peak, the convergence scale decreases to $\sim 300$\,pc, and it increases again as the SFR decreases during the Plateau stage. Finally, the convergence point shifts to $\sim100$\,pc in the most intense starburst during the second SF peak. This corresponds to the average separation between the peaks of molecular gas and SF shrinking as the SFR increases. The opening of the tuning fork at the smallest scale also decreases during the merger, reflecting that the contrast of spatial de-correlation between recent SFR and dense gas decreases.

When two galaxies collide, the projection effect could be substantial: several star-forming regions may appear to overlap along the line-of-sight, reducing both the convergence scale and opening. To justify our findings in the 2D tuning fork, we extended the traditional 2D tuning fork analysis to 3D space in the next subsection.

\subsection{3D tuning fork diagram}
\label{sec:3dtf}
To generate the 3D tuning fork diagram, we first used a cloud-in-cell (CIC) algorithm to assign the particle data to a Cartesian grid. Consistent with the 2D version, we only selected the gas particles with $2x_\text{\ce{H2}}>0.1$ for the molecular gas cube and 2--5\,Myr age young stars for the SFR cube. We smoothed the cubes using a 3D Gaussian filter with a width of 20\,pc and identify the local extrema of $\rho_\text{\ce{H2}}$ and $\dot{\rho}_\star$ in the smoothed cubes. We then smoothed the cubes with a top hat filter and compute the average $\rho_\text{\ce{H2}}$ and $\dot{\rho}_\star$ at the locations of gas and SFR peaks. The resulting 3D tuning fork diagram is presented in the bottom panel of Fig.~\ref{fig:tuingfork}.

Similar to the 2D tuning fork diagram, the convergence point moves toward a smaller scale during the ongoing merger, reaching $\sim100$\,pc during the most intense starburst at the second SF peak. However, the opening of the tuning fork does not decrease as evidently as in the 2D case.
Therefore, this 3D tuning fork diagram confirms that the merger does, in fact, decrease the average separation between regions of SF and dense clouds, whereas the reduction of the opening of the two branches is possibly a projection effect.

The reduced spatial de-correlation between young stars and dense clouds observed in the simulated merger can be explained by various physical causes. This might indicate that the continued accumulation of dense gas and high environmental pressure during a merger reduced the ability of stellar feedback to disperse clouds. Alternatively, stellar feedback may remain largely unchanged at the cloud scale, but the volume density of individual clouds at kiloparsec scale increases. Our findings in Section~\ref{sec:cloud-scale} favor the latter scenario, highlighting the importance of early stellar feedback. Future simulations of larger systems can further examine this finding and synergy with the ALMA observations of nearby mergers.

\section{Discussions and conclusions}
\label{sec:discuss}

\subsection{Energy of compressive tides}
\label{sec:tidal-energy}
In Section~\ref{sec:cloud-scale}, we found that the strong tidal fields present during a merger boost the accumulation of cold clouds; however, they do not alter the SFE of these clouds that are forming massive clusters. Here, we present some order of magnitude estimates to illustrate these phenomena from an energy-based perspective.

Assuming an isotropic tidal field, the tidal energy of a spherical cloud with mass $M$ and radius $R$ can be estimated as  
\begin{align}
  E_\text{tid}&=\frac{1}{2}\lambda_\text{tid}MR^2\notag\\
  &=10^{46}\,\text{erg}\left(\frac{\lambda_\text{tid}}{10^4\,\text{Gyr}^{-2}}\right)\left(\frac{M}{10^3\,\Msun}\right)\left(\frac{R}{10\,\text{pc}}\right)^2\,.
\end{align}

This is much smaller than the energy released by a single SN explosion ($10^{51}$\,erg), and even the early stellar wind and radiation feedback \citep[e.g.,][]{2013ApJ...770...25A,2021MNRAS.505.3470J}. Thus, even a single massive star can dominate the evolution of the clouds.

However, the binding energy of such a cloud is only
\begin{align}
  U_\text{bind}&=-\frac{3}{5}\frac{GM^2}{R}\notag\\
  &=-5\times10^{45}\,\text{erg}\left(\frac{M}{10^3\,\Msun}\right)^2\left(\frac{R}{10\,\text{pc}}\right)^{-1}\,.
\end{align}
Therefore, the tidal energy is actually comparable to the binding energy of a cloud.
If there is no massive star present in a cloud, the SFE will strongly depend on the tidal strength, as seen in Section~\ref{sec:small-clouds}.

In Section~\ref{sec:small-clouds}, we found several clouds in strong tidal fields with high integrated SFE, and we suggested that they are compact enough to be free from tidal disruption. To quantify this, we compare the tidal energy and binding energy by
\begin{equation}
\left|\frac{U_\text{bind}}{E_\text{tid}} \right|\sim \frac{GM}{\lambda_\text{tid}R^3}\sim \frac{\beta G\bar{\rho}}{\lambda_\text{tid}}\,,
\end{equation}
where $\bar{\rho}$ is the average density of the cloud and $\beta$ is a geometric factor.

Given that the maximum tidal strength in our simulation is $\lambda_\text{tid}\sim10^5$\,Gyr$^{-2}$, we obtain the critical density of  $\bar{\rho}_\text{crit} = 22\beta\,\Msun\,\text{pc}^{-3}$ for ${u_\text{bind}}/{u_\text{tid}}=1$. Clouds with lower average density should be susceptible to the tidal effects while much denser clouds are tightly bound. In our simulation, we found that the actual volume density of the outlier clouds in Fig.\ref{fig:eps_tidal} (i.e., clouds that have SFE within strong tidal fields) exceeds  $200\,\Msun$\,pc$^{-3}$, consistent with the above estimate.

On the other hand, at a large scale, the tidal energy can far exceed the self-gravitational energy. In the warm neutral medium (WIM), which dominates the ISM in dwarf galaxies, the $p{\rm d}V$ work done by tidal force can be expressed as 
\begin{equation}
  {\rm d}U_\text{int}=4\pi nkTR^2{\rm d}R\,,\notag
\end{equation}
and the change of tidal energy is 
\begin{equation}
 {\rm d} E_\text{tid}=\lambda_\text{tid}MR{\rm d}R\,.\notag
\end{equation}
Combining these two equations, we can obtain a characteristic scale analogous to the Jeans length:
\begin{align}
  R_\text{tid}=\sqrt{\frac{3\pi kT}{\lambda_\text{tid}\mu m_\text{H}}}=280\,\text{pc}\left(\frac{\lambda_\text{tid}}{10^4\,\text{Gyr}^{-2}}\right)^{-0.5}\left(\frac{T}{10^4\,\text{K}}\right)^{0.5}\,.
\end{align}
As a comparison, the Jeans length in $T=10^{4}$\,K, $n_\text{H}=1$\,cm$^{-3}$ WIM is $1.6$\,kpc, much larger than $R_\text{tid}$. In the isolated dwarf galaxies or solar neighborhood tidal field with $\lambda_\text{tid}=1600$\,Gyr$^{-2}$, $R_\text{tid}$ is comparable to the Jeans length, while in the strong merger tidal field with $\lambda_\text{tid}>10^5$\,Gyr$^{-2}$, $R_\text{tid}$ decreases to $<100$\,pc. Therefore, the compressive tides can lead to gravitational instability at a much smaller scale compared with self-gravity and facilitate the formation of dense star-forming clouds. In the case of our simulation, we found an enhanced cloud formation during the merger with a mean separation of tens to hundreds of parsecs.

\subsection{Comparison with other work}
\label{sec:comp}

As the IC of our simulation is the same as \citetalias{2020ApJ...891....2L}, we can make a direct comparison with the series of GRIFFIN papers. The morphology and SFR of our merger system show good agreement with \citetalias{2020ApJ...891....2L}. The slope of their CIMFs varies between $-1.7$ and $-2$ during the merger, comparable but slightly flatter than our result of $-1.98$. More importantly, the most massive cluster in \citetalias{2020ApJ...891....2L} reaches a mass of $7\times10^5\,\Msun$, while we only obtain $\sim3\times10^4\,\Msun$ clusters. Thus, our model lacks massive clusters compared to \citetalias{2020ApJ...891....2L}.
As \citetalias{2020ApJ...891....2L} we used the same algorithm to find young clusters (though they define young as $<10$\,Myr while our definition is $<5$\,Myr), the difference in CIMF and maximum cluster mass should be attributed to the SF and feedback model. We notice that the present observation cannot constrain the models with such details [the only hint is that the maximum cluster mass of $7\times10^5\,\Msun$ in \citetalias{2020ApJ...891....2L} seems too large for their peak SFR of $\sim0.2\,\Msun$\,yr$^{-1}$ (see Fig.\ref{fig:SFR-Mcl})], yet it is valuable for a theoretical exploration. 

The SF model determines the site, timing, and dynamics of newly formed stars, thus determining the properties of clusters in terms of the mass, size, boundness of clusters and the CIMF \citep{2022MNRAS.509.5938H}. As both \citetalias{2020ApJ...891....2L} and our model do not resolve the formation dynamics of stars, the size and boundness of clusters can be strongly affected by the underlying SF density criterion and the gravitational softening. Since the gas mass resolution of our simulation is higher, we adopted a relatively high density criterion of $3000$\,cm$^{-3}$ and a small softening for stars of $0.05$\,pc. As a comparison, the Jeans criterion adopted by \citetalias{2020ApJ...891....2L} typically allows star formation at $500$\,cm$^{-3}$, with a softening of $0.1$\,pc. Therefore, the clusters in our simulation are more compact, especially for the small cluster formed through a monolithic cloud collapse. Such compact clusters are thus hard for large clusters to accrete and assemble. More sophisticated formation and dynamics models \citep[e.g.,][]{2023MNRAS.522.3092L,2024MNRAS.530..645L,2025MNRAS.538.2129L:Lahen} are necessary to generate more realistic cluster populations.

\begin{figure}
	\includegraphics[width=\columnwidth]{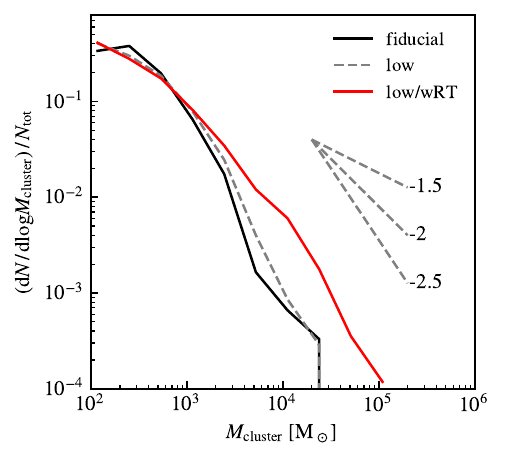}
    \caption{CIMFs for the simulation with weakened radiative feedback. The red curve shows the result of $10\,\Msun$ resolution stellar radiation luminosities reduced by a factor of 0.1 across all bands. Solid black and dashed gray curves represent the results from the simulation with normal radiative feedback at $2\,\Msun$ and $10\,\Msun$, respectively. The CIMF is not sensitive to the numerical resolution but is very susceptible to the intensity of radiative feedback.
    }
    \label{fig:wRT}
\end{figure}

Another major difference between our model and \citetalias{2020ApJ...891....2L} is that we used radiative transfer in \areport to explicitly model the radiative feedback from massive stars, while \citetalias{2020ApJ...891....2L} adopted a Str\"omgren-type approximation method to model the effects of radiation. \cite{2017MNRAS.471.2151H} noticed that their approximation method preferentially ionized mass concentration regions with high density. This can potentially reduce the momentum injection rate by \HII region expansion \citep[e.g.,][]{2002ApJ...566..302M:Matzner,2024MNRAS.527..478D}. Thus, it can be interesting to test the dependence of CIMF on the intensity of radiative feedback. We ran a low-resolution simulation using $10\,\Msun$ resolution and stellar radiation luminosities reduced by a factor of 0.1 across all bands, labeled as “low/wRT” (weak radiative feedback). As a controlled group, we also ran a $10\,\Msun$ resolution simulation with normal luminosities. We show the CIMFs of these simulations in Fig.~\ref{fig:wRT}. The CIMF is not sensitive to the numerical resolution, but is very susceptible to the intensity of radiative feedback. The weaker radiative feedback leads to a flatter slope and the most massive cluster reaches a mass of $>10^5\,\Msun$. This result suggests that the process of cluster formation is influenced not merely by the presence of early feedback, but is also sensitive to the strength of this feedback, requiring an accurate modeling of all stellar feedback channels.

\cite{2024MNRAS.534..215F:Fotopoulou} studied the cold clouds in a similar dwarf-dwarf merger simulation. Their cloud selection criteria are slightly different from ours: they did not use the sub-virial criterion, and they adopted the same density criterion as ours but an additional temperature cut of $T<100$\,K. 
Contrary to \cite{2024MNRAS.534..215F:Fotopoulou}, we find that the slope of the cloud mass function varies significantly across the merger stages. Although it can be related to the cloud selection criterion, it is also important to study whether the cloud mass function is sensitive to the SF and feedback model in the future.
Additionally, they find that the cloud lifetimes do not seem to be affected by the environment. Our results confirm this finding and explain it by the feedback-dominated cloud life cycle. They find that the clouds have an e-folding lifetime distribution with a characteristic timescale exceeding $3$\,Myr, potentially due to not restricting the virial status of the clouds, which allows them to encompass a longer period of accretion and collapse.

\cite{2022MNRAS.514..265L:Li} studied the major merger of two MW-like galaxies with similar orbital configuration using the SMUGGLE model \citep{2019MNRAS.489.4233M}. Similar to our results, they also find that both the mass function of GMCs and young clusters vary significantly across the merger stages. The slope of the GMC mass function varies from 2.47 in isolated galaxies to 1.99 in mergers, and the CIMF slope varies from 1.93 to 1.67. Since the MW-like galaxies are much more massive, the tidal strength in their simulation extends to $\lambda_\text{tid}\gtrsim10^7$\,Gyr$^{-2}$, and they find a clear positive relation between tidal field strength and young cluster mass in this large $\lambda_\text{tid}$ range. Since we find the cloud-scale SFE to be insensitive to the tidal environment, this cluster-tide correlation can be a direct result of the positive correlation between $\lambda_\text{tid}$ and $M_\text{cloud}$ presented in Fig.~\ref{fig:Mmax_tidal}. The SFR in the MW merger is enhanced by a factor of 10. This enhancement is an order of magnitude weaker compared to that observed in our dwarf-dwarf merger, whereas their galaxy-wide SFE, conversely, is comparable during the peak SF phases and even presents a roughly factor of 2 higher than the maximum value. This suggests that major mergers can exert a more significant influence on SF in low-mass galaxies. We can understand this by considering that SF is less efficient in isolated dwarf galaxies compared to the MW disk. This inefficiency arises because the dwarf disk is predominantly composed of warm, diffuse HI gas. In contrast, during merger processes, this diffuse HI gas is accumulated into the central region, leading to the formation of star-forming dense clouds.

\cite{2025A&A...699A.282N:Ni} studied the evolution of giant molecular clouds (GMCs) in a suite of MW-mass galaxies. They find that the integrated SFE of the GMCs can also be described by a $\epsilon_\text{int}$--$\Sigma_0$ relation. They also notice that the galactic dynamics can introduce scatter in this relation. Our finding for the feedback-free clouds in Section~\ref{sec:small-clouds} provides an extreme example for this: the SFE of clouds without massive stars is strongly correlated with the intensity of tidal fields.

Recently, \cite{2025arXiv250301949S:Shen} and \cite{2025arXiv250505554W:Wang} studied the cloud-scale SFE in the {\sc thesan-zoom} cosmological simulation of high-redshift galaxies \citep{2025arXiv250220437K:Kannan}. Although they did not track the evolution of individual GMCs, they find that GMCs exhibit universal statistics in terms of mass functions, size, surface density, and especially cloud-scale SFE, regardless of the environments. The feedback-regulated cloud-scale SFE identified in this work offers a direct explanation for their results, suggesting that our findings might be applicable to describe the SF cycles throughout cosmic time.

Previous numerical experiments have demonstrated that the opening and convergence points of tuning fork diagrams are notably influenced by the feedback model, particularly the pre-SN feedback \citep{2019MNRAS.487.1717F,2021ApJ...918...13S:Semenov}. \cite{2020MNRAS.498..385J:Jeffreson} further proposed that they can be affected by galactic dynamics in MW-like galaxies. Our finding in Section~\ref{sec:tuningfork} underscores the importance of galactic dynamics and predicts a significantly reduced spatial de-correlation in merger systems.

\subsection{Limitations of this work}
We note that there are a few limitations in this work. First, we used a highly idealized prograde major merger of two identical dwarf galaxies, which provides the maximum effects of tidal interaction and highest enhancement of gas accumulation. A comprehensive study requires a thorough exploration of the properties of host galaxies and orbit parameters. Moreover, due to the low mass of the dwarf galaxies, the boost of the tidal field during the merger is within two orders of magnitudes, and the strongest tides only reach $\lambda_\text{tid}\approx10^5$\,Gyr$^{-2}$. The major merger of MW-mass galaxies can further boost the tidal strength to $\lambda_\text{tid}>10^7$\,Gyr$^{-2}$ \citep[e.g.,][]{2022MNRAS.514..265L:Li}, which potentially results in more drastic effects on the cloud evolution. The gas surface density and the SFR surface density of our system are also too low to directly compare with spatially resolved observations of nearby massive galaxies. As discussed in Section~\ref{sec:comp}, our model is not yet able to accurately capture the internal structure of star clusters and, therefore, cannot investigate their long-term dynamical evolution. The time and location of massive star formation do not have any preference in our model, which may be an oversimplification. For example, if massive stars preferentially form in denser cores and at later times compared with low-mass stars \citep{2022MNRAS.512..216G,2024MNRAS.527.6732F}, the cloud-scale SFE can be higher with respect to our prediction. The compact cluster formation could also be attributed to this oversimplified SF model. Additionally, the resolved collisional dynamics might considerably expand the cluster sizes, driven by stellar dynamical processes \citep{2025MNRAS.538.2129L:Lahen}. We are working on reproducing a more realistic star cluster population in the RIGEL model (Deng et al. in prep.).

\subsection{Summary and conclusions}
In this paper, we attempted to find the mechanisms that drive starbursts in dwarf--dwarf major mergers using a high-resolution numerical simulation. Our key findings are summarized as follows:

First, we find that high-pressure, dense gas significantly increases due to angular momentum cancellation and tidal compression. The total mass of $n_\text{H}>100$\,cm$^{-3}$ dense gas increased by a factor of 56 during the merger. 

However, the depletion time of dense star-forming gas, especially the gas in clouds, remains roughly constant across all merger stages. The depletion time of gas in clouds fluctuates around its median value of 7.4\,Myr, with the 16--84 percentile range of 4.7–12\,Myr. This suggests that the merger does not significantly change the cloud-scale SFE. 

Indeed, by tracking the evolution of individual clouds, we find that the lifetimes of clouds across all stages follow the same exponentially decaying distribution with a characteristic timescale of $\tau_\text{life}=1.48$\,Myr. Moreover, the integrated SFE of clouds that formed at least one massive star ($M_{\star,\text{final}}>100\,\Msun$) can always be described by a $\epsilon_\text{int}$--$\Sigma_0$ relation assuming the feedback-gravity balance. The median integrated SFE during the most intense starburst was lower by only 0.17–0.33 dex compared to that in isolated galaxies. This suggests that the SF activities in the merger system are regulated by stellar feedback rather than galactic dynamics. The integrated SFE in clouds without massive stars presents a strong correlation with tidal strength, while these clouds only contribute 10\% of their total stellar mass.

The compressive tidal fields in the merger system play an essential role in compressing and accumulating gas to form cold dense clouds. The cloud mass is positively correlated to the tidal strength, and massive clouds preferentially reside in strong tidal fields. As a result, the slope of the cloud mass function becomes flatter from $-2.36$ upon reaching $-1.79$ in the $2^\text{nd}$ peak. Larger clouds serve as locations for the formation of massive clusters. Indeed, we also notice an alteration in the slope of the CIMF from $-2.23$ to $-2.03$, even though this variation is less pronounced compared to the change in the cloud mass function. 

The average separation of clouds shows a strong negative correlation with tidal strength: $\lambda_\text{tid}\propto l^{-0.43}$. It is compressed from hundreds of parsecs in isolated galaxies to $<50$\,pc during the merger. As a result, the spatial de-correlation between clouds and clusters observed in the tuning fork diagram also decreases from $\gtrsim1$\,kpc to $\sim0.1$\,kpc during the merger.

Based on these findings, we can conclude that the merger stimulates the formation of dense clouds fueling the SF due to the strong compressive tidal fields. However, once massive stars form in the clouds, they determine the further evolution of their natal clouds and eventually the integrated SFE. Therefore, the cloud-scale SFE or depletion time remains unchanged during the merger.

\begin{acknowledgements}
We thank Volker Springel for giving us access to \arepo, and thank Natalia Lah\'en for helping us generate the initial condition. YD is grateful to Zhiyuan Li, Zhiqiang Yan, and Yulong Gao for useful discussions. 
HL is supported by the National Key R\&D Program of China No. 2023YFB3002502, the National Natural Science Foundation of China under No. 12373006, and the China Manned Space Program with grant No. CMS-CSST-2025-A10.
RK acknowledges support of the Natural Sciences and Engineering Research Council of Canada (NSERC) through a Discovery Grant and a Discovery Launch Supplement (funding reference numbers RGPIN-2024-06222 and DGECR-2024-00144) and York University's Global Research Excellence Initiative. BL acknowledges the funding of the Deutsche Forschungsgemeinschaft (DFG, German Research Foundation) under Germany's Excellence Strategy EXC 2181/1 - 390900948 (the Heidelberg STRUCTURES Excellence Cluster). We use \textsc{python} packages {\sc NumPy} \citep{harris2020array}, {\sc SciPy} \citep{2020SciPy-NMeth}, {\sc astropy} \citep{2013A&A...558A..33A,2018AJ....156..123A}, {\sc matplotlib} \citep{Hunter:2007}, and {\sc Paicos} \citep{2024JOSS....9.6296B:Berlok} to analyze and visualize the simulation data.
\end{acknowledgements}

\bibliographystyle{aa}
\bibliography{aa56854-25}

\begin{appendix}
\end{appendix}
\end{document}